\documentclass[11pt]{article}
\usepackage{amsfonts,amsbsy,bm,euscript,mathrsfs}
\usepackage{amssymb,stmaryrd,faktor}
\usepackage{amsmath}
\usepackage{bbm}
\usepackage{graphicx}
\usepackage[title,titletoc]{appendix}
\usepackage[bookmarks=true,colorlinks=true,linkcolor=blue,citecolor=blue,urlcolor=blue,bookmarksnumbered]{hyperref}
\usepackage{dsfont}
\usepackage{lmodern}
\usepackage{mathrsfs}
\usepackage{mathtools}
\usepackage{bbm}
\usepackage{braket}
\usepackage{slashed}
\usepackage[normalem]{ulem}
\usepackage{booktabs}
\usepackage{tikz}
\usepackage[utf8]{inputenc}
\usepackage[english]{babel}
\usepackage{a4wide}
\usepackage{cite}
\usepackage{appendix}
\usepackage{etoolbox}
\usepackage{physics}
\usepackage{sectsty}
\usepackage{float}
\usepackage{multirow}
\usepackage{systeme}
\usepackage{xcolor}
\usepackage{mathdots}
\usepackage{nicematrix}
\usepackage{subcaption}

\DeclareMathOperator{\e}{e}
\DeclareMathOperator{\im}{i\!}

\DeclareMathOperator{\Cc}{\mathit{C}}

\newcommand\blfootnote[1]{
  \begingroup
  \renewcommand\thefootnote{}\footnote{#1}
  \addtocounter{footnote}{-1}
  \endgroup
}

\def\texto
       {\topmargin -20pt
        \oddsidemargin 0pt
        \headheight 0pt
        \headsep 0pt
        \textwidth 6.35in
        \textheight 9.25in
        }
\texto

\def\baselinestretch{1.2}
\sectionfont{\large}
\renewcommand{\theequation}{\thesection.\arabic{equation}}
\csname @addtoreset\endcsname{equation}{section}

\begin{document} 

\begin{titlepage}

\begin{center}

\phantom{\LARGE X}

\vskip 0.25in

{

\large\textbf{
Efficient Eigenstate Preparation in an Integrable Model \\[1mm]
with Hilbert Space Fragmentation
}
}

\vskip 0.45in

\textbf{
Roberto Ruiz\footnote{{\texttt{roberto.ruiz@ift.csic.es}}}$^{a\dagger}$,
Alejandro Sopena\footnote{{\texttt{alejandro.sopena@tii.ae}}}$^{b,a\dagger}$,
Balázs Pozsgay\footnote{{\texttt{pozsgay.balazs@ttk.elte.hu}}}$^{c}$,
Esperanza L{\'o}pez\footnote{{\texttt{{\texttt{esperanza.lopez@csic.es}}}}}$^{a}$
\setcounter{footnote}{0}
\blfootnote{\hspace{-4.425pt}$\vphantom{x}^\dagger$The first two authors contributed equally.} 
}

\vskip 0.15in

$\vphantom{x}^{a}$\! Instituto de F{\'i}sica Te{\'o}rica UAM/CSIC \\
Universidad Aut{\'o}noma de Madrid \\
C/ Nicolás Cabrera 13--15 \\
Cantoblanco, 28049 Madrid, Spain \\ 

\vskip 0.15in

$\vphantom{x}^{b}$\! 
Quantum Research Center\\ 
Technology Innovation Institute\\ 
Abu Dhabi, UAE \\

\vskip 0.15in

$\vphantom{x}^{c}$\! 
MTA-ELTE “Momentum” Integrable Quantum Dynamics Research Group,\\
ELTE Eötvös Loránd University\\
Pázmány Péter sétány 1/a,\\
1117
Budapest, Hungary \\

\vskip 0.15in

\vskip 0.1in 

\end{center}

\vskip .5in

\noindent 

\vskip .1in

\noindent 
We demonstrate that eigenstates of selected interacting spin chains can be efficiently prepared using quantum circuits with polynomial depth in system size and particle number. While this is well established for free-fermionic spin chains, we extend this result to the folded XXZ model,  
an integrable rigid-rod deformation of the XX model with simple interactions that exhibits Hilbert space fragmentation.  
We construct explicit quantum circuits that efficiently prepare arbitrary eigenstates of this model on an open chain. 
Error-mitigated noisy simulations with up to 13 qubits and various qubit connectivities achieve a relative error below 5\%. 
As a byproduct, we extend a recent reformulation of the Bethe Ansatz as a quantum circuit to open boundary conditions.

\begin{sloppypar}

\noindent

\end{sloppypar}

\vskip .4in

\vskip .1in

\noindent

\end{titlepage} 

\def\baselinestretch{1}
\baselineskip 15pt
\sectionfont{\large} 
\renewcommand{\theequation}{\thesection.\arabic{equation}} 
\csname @addtoreset\endcsname{equation}{section}

\section{Introduction}

Eigenstate preparation is an important challenge in the application of quantum computation to the study of many-body physics.
The implementation of eigenstates on a quantum computer would give direct access to their non-local properties, which are notably hard to obtain via classical computations.
Moreover, eigenstates can be useful for benchmarking as well as for the initialization of quantum algorithms. 

Quantum circuits for eigenstate preparation must be efficient to be implementable, meaning
that the number of one- and two-qubit gates must grow polynomially with the number of qubits and/or excitations in the model.
There are efficient circuits that prepare free-fermion eigenstates exactly~\cite{Verstraete09,Jiang17,Kivlichan18,CerveraLierta18,Farreras24}.
The existence of such algorithms relates to the fact free fermions are
classically simulatable~\cite{Valiant02,Terhal02,Jozsa08}.
What is the class of models that admit efficient eigenstate preparation with unitary circuits remains an open question.
It is true that ground states and certain excited states can have a relatively simple structure that enables efficient preparation.
Conversely, our goal here is to develop algorithms capable of preparing arbitrary excited states in selected models.

It is natural to consider integrable spin chains with regard to this problem. 
Integrable systems are reputed to have simpler eigenstates than generic models, 
as integrability guarantees the models have completely elastic and factorized scattering processes~\cite{sutherland-book}. 
The exact wavefunctions, 
which are computed by the coordinate Bethe Ansatz (CBA), 
are often known. 
The associated eigenstates can be usually constructed efficiently by a sequence of non-unitary linear operations, 
in a procedure known as algebraic Bethe Ansatz (ABA)~\cite{Korepin1993kvr,faddeev1996algebraic}. However, it is not known how to prepare the eigenstates efficiently using unitary quantum circuits.

The main motivation for studying this problem is the measurement of local and non-local correlation functions in these states, which could be used to benchmark quantum computers and demonstrate quantum supremacy. We now elaborate on these motivations in detail.

In integrable spin chains there are many analytic results available concerning the correlation functions in the Bethe states. 
Consider for example the XXZ Heisenberg spin chain, which is a paradigmatic integrable model. Mean values of local operators are known in arbitrary excited states \cite{sajat-corr2}. This could be used in benchmarking quantum computations, by preparing selected eigenstates in small system sizes, and measuring their local correlations, for example diagonal $\langle \sigma^z_1 \sigma^z_j \rangle$ correlation functions. Such an approach works well until the measured two-point correlation function is sufficiently local. The known exact results of \cite{sajat-corr2} are based on the theory of factorized correlation functions (see for example \cite{kluemper-goehmann-finiteT-review}), which involve certain algebraic manipulations that are tractable on a classical computer as long as the distance is not larger than roughly 8-9 sites. For longer distances the complexity becomes too large and the classical computation becomes infeasible. On the other hand, if there were efficient quantum circuits that prepare the eigenstates on a quantum computer, then one could compute correlation functions 
with distance roughly greather than $10$, and this would be a demonstration of quantum supremacy. The intermediate range correlation functions are the only ones that are not accessible to classical computations, because the asymptotic limit is often known, even in cases of algebraically decaying correlation functions in the ground state of the massless regime \cite{karol-hab}. There are also numerical methods based on matrix-product states (MPS), which can be used to numerically compute correlation functions on larger spin chains. However, such methods are restricted to ground states, or more generally to eigenstates with sub-volume law entanglement. In contrast, excited states typically have volume law entanglement~\cite{bianchi_volume-law_2022}.

Following these motivations, quantum circuits for eigenstate preparation have attracted considerable attention in the recent years.
Quantum circuits to prepare the eigenstates of the XXZ model with closed~\cite{VanDyke21,Sopena22,Ruiz23, Raveh24} 
and open boundary conditions~\cite{VanDyke22,Raveh24}
have been designed.
They fall into two classes: probabilistic~\cite{VanDyke21,VanDyke22}
and deterministic algorithms~\cite{Sopena22,Ruiz23,Raveh24}.
The number of gates of the probabilistic algorithms in~\cite{VanDyke21,VanDyke22} is polynomial in the number of qubits $N$ and excitations (``magnons'') $M$, but
their success probability is exponentially small~\cite{Li22}. 
The number of gates of the deterministic algorithm of~\cite{Raveh24} is exponential in the number of magnons. 
Moreover, the deterministic circuits of~\cite{Sopena22,Ruiz23} involve multi-qubit unitaries, whose decomposition in terms of elementary gates likely suffers from the same problem.
There are thus no known efficient quantum algorithm preparing the XXZ eigenstates. 
This raises the question of whether a quantum computer can access general eigenstates beyond those of free fermions. Our main result is proving that there are
interacting many-body systems for which efficient 
algorithms prepare all their eigenstates.

Our strategy is to consider interacting models with relatively simple dynamical features. Specifically,
we focus on the strong-coupling limit of the XXZ spin chain, where the dynamics drastically simplifies.
There are essentially two ways to define an infinite-coupling limit of this spin chain,  
both of which can be understood as Schrieffer-Wolff transformation of certain kind. 
First, 
following~\cite{xxz-triple-point}, 
the model can be scaled to its triple point, 
which brings about the
``constrained XX model''.
The interactions of this spin chain consist of contact repulsion among magnons. 
The related
but slightly different strong-coupling limit 
of~\cite{Zadnik20,Zadnik21,Pozsgay21} 
leads to the ``folded XXZ model''~\cite{folded-XXZ-stochastic-1,Yang19},   
whose excitations are magnons and domain walls. The Hilbert space of this spin chain splits into  exponentially many subsectors with disconnected dynamics, which display the phenomenon called ``Hilbert space 
fragmentation''~\cite{fragmentation-scars-review-2,hfrag-review}. This phenomenon, called ``Hilbert space 
fragmentation'', has attracted recently much attention because it provides a novel form of weak ergodicity breaking~\cite{fragmentation-scars-review-2,hfrag-review}.
The folded XXZ model coincides with the constrained XX model in the purely magnonic sector.
The simplicity of the folded XXZ model enabled exact computations in a variety of 
situations~\cite{Pozsgay21,folded-jammed,folded4}. 

In this paper, we construct a quantum circuit capable of preparing arbitrary
eigenstates of the folded XXZ model with open boundaries exactly. The number of gates grows quadratically with the number of sites and linearly with that of excitations, being thus efficient. The structure of the circuit is the following.
The initial state fed to the circuit is an $M$-magnon eigenstate of the free XX model. 
The first step is the application of a unitary $U_0$, which implements contact repulsion in such a way that the initial state is mapped onto an eigenstate of the constrained XX model. The output state then undergoes a second unitary $V_D$ which introduced the domain walls. Both $U_0$ and $V_D$ are built out of $\mathrm{CNOT}$ and $\mathrm{CSWAP}$. For the preparation of the XX eigenstates, we generalize the quantum algorithm of~\cite{Sopena22,Ruiz23}, which prepare Bethe states on periodic chains, to open boundary conditions. We also perform numerical simulations of our circuits for several eigenstates with few excitations and sites. We consider different qubit connectivities and represent noise through a depolarizing channel.

The paper is organized as follows.
In Section~\ref{sFolded}, we review the main properties of the folded XXZ model, 
explaining how Hilbert space fragmentation arises. 
In Section~\ref{sec:eigenstates},
we present the Bethe eigenstates of the model. 
In Section~\ref{squantalg},
we construct a quantum circuit
whose input are eigenstates of the XX model and output those of the folded XXZ chain. 
In Section~\ref{ssim}, we  present
numerical simulations of our quantum circuit.
In 
Section
\ref{sQA_other_models},
we present other models and states to which a simple generalization of our quantum algorithm would possibly apply.
In Section~\ref{sconcl}, we conclude with general remarks and prospects
on future research.
Appendices~\ref{appOABC}--\ref{appcompilation} contain technical details.

\section{The Folded XXZ Model}

\label{sFolded}

In this section, we introduce the {folded XXZ model} and discuss its main properties. 
The model is a spin-$1/2$ chain whose Hamiltonian is
\begin{equation}
  H=-\frac{1}{8}\sum_{j}(1+\sigma^z_{j}\sigma^z_{j+3})(\sigma^x_{j+1} \sigma^x_{j+2}+\sigma^y_{j+1} \sigma^y_{j+2}) \ .
  \label{foldedH}
\end{equation}
The folded XXZ model is one of the simplest interacting spin chains. The model is integrable: it has an infinite family
of local conserved charges. Furthermore, the model displays Hilbert space fragmentation.
We discuss these  properties below. We stress that there are models for discrete time evolution based on the folded XXZ model~\cite{sajat-cellaut} that have been used in the analysis anomalous transport~\cite{vasseur-anom}, but in this paper we consider only the Hamiltonian~\eqref{foldedH} and not the associated cellular automata. 

The Hamiltonian density 
describes kinetically constrained hopping processes. 
If we identify
spin-up and spin-down states
with the computational basis of the qubit like
\begin{equation}
    \ket{\uparrow} := \ket{0}
\ , \quad 
    \ket{\downarrow} := \ket{1}
  \ ,
\end{equation}
the processes allowed by~\eqref{foldedH} are
\begin{equation}
  \ket{0100} \ \leftrightarrow \ \ket{0010} \ , \quad
    \ket{1011} \ \leftrightarrow \ \ket{1101} \ .
    \label{trans}
\end{equation}
We often use qubit rather than spin nomenclature.
The Hamiltonian~\eqref{foldedH} clearly
has two global $U(1)$ symmetries.
The first corresponds to the charge
\begin{equation}
  Q_1=\frac{1}{2}\sum_j (1-\sigma^z_j) \ ,
\end{equation}
which counts the number 
qubits at $\ket{1}$.
The second corresponds to the charge 
\begin{equation}
  Q_2=\frac{1}{2}\sum_j (1-\sigma^z_j \sigma^z_{j+1}) \ ,
\end{equation}
which counts
the number of 
changes from zero to one in a bit string.
While both the Hamiltonian and $Q_2$ are invariant under a global bit flip.
$Q_1\mapsto N/2-Q_1$ under this transformation instead. 

The folded XXZ model can be derived from a special strong-coupling limit of the standard XXZ model with coupling constant 
$\Delta$, whose Hamiltonian is
\begin{equation}
    \label{HXXZ}
  H_{\mathrm{XXZ}}=\sum_j  (\sigma^x_{j}\sigma^x_{j+1}+\sigma^y_{j}\sigma^y_{j+1}+\Delta \sigma^z_{j}\sigma^z_{j+1}) \ .
\end{equation}
The Hamiltonian~\eqref{foldedH}
follows from both
a Schrieffer-Wolff transformation at $\Delta\rightarrow\infty$~\cite{Zadnik20} and
the $\Delta\rightarrow\infty$ limit of the
fourth element of the family of local conserved charges
of the Hamiltonian~\eqref{HXXZ}~\cite{Pozsgay21}.
Both approaches are
regular at the level
of the eigenstates. 
Therefore, all
the eigenstates of the folded XXZ model match a $\Delta\to\infty$
limit of the eigenstates of the standard XXZ model.
Furthermore, the existence of an infinite
family of higher conserved charges follows from the
$\Delta\to\infty$ limit of the family of the XXZ model~\cite{Pozsgay21}.

The $\Delta\to\infty$ limit cannot be applied
directly to the algebraic
objects guaranteeing integrability.
The Lax operators, the transfer matrices, and the
R-matrices of 
the XXZ model are singular when $\Delta\to\infty$.
New objects are needed to prove integrability in the
framework of the Yang-Baxter equation~\cite{sajat-medium}. 
The impediment implies that the construction of the eigenstates by the ABA can not be applied to the folded XXZ model directly. Furthermore, there is no obvious generalization.

However, it is possible to find all eigenstates 
of the folded XXZ model by means of
the standard CBA, which uses linear superpositions of plane waves as 
trial functions. The applicability of the CBA
was shown in the
alternative approaches of~\cite{Zadnik20}
and~\cite{Pozsgay21}, the latter of which we follow.
Before we review the wavefunctions of the CBA in Section~\ref{sec:eigenstates},
we present the physics of the spin chain in this section. 
In Subsection~\ref{sec:dynamics}, we describe the
dynamical processes of the model. We then address
the mechanism of Hilbert space fragmentation in Subsection~\ref{sec:fragm}.

\subsection{The Dynamics of the Model}

\label{sec:dynamics}

Physical processes of the folded XXZ model
can be understood
as the dynamics of isolated particles
and holes together with domain walls.
We define these concepts 
with the help of the computational
basis first. We analyze transitions between states afterward.

We say that a transition
is a link between two qubits
such that two neighboring qubits carry different bits.
Examples of states with transitions
are
$\ket{01}$ and $\ket{10}$.
A particle is a single $\ket{1}$ embedded into a block $\ket{00\ldots0}$,
and a hole is a single $\ket{0}$ embedded into a block $\ket{11\ldots 1}$.
For instance, $\ket{00100}$ has a particle at the third position
and $\ket{11011}$ a hole
at the third position.
We use the common term ``magnon'' for both particles and holes.
Magnons can be
seen as a bound state of two transitions at neighboring positions.
The state $\ket{0010011011}$ thus 
has two magnons: a particle on the third position
and a hole on the eight position.
Our definition implies that magnons cannot occupy neighboring positions, but
they can occupy next-to-nearest neighboring positions.
The position of magnons can be ambiguous in certain states.
For example, we can say
$\ket{1010}$ has two particles at the first
and third position
or two holes at the second
and fourth positions. 
We resolve the ambiguity by 
choosing magnons to be at
leftmost available configuration, hence 
$\ket{1010}$
has two particles rather than two holes.

A domain is a block of the
spin chain with more than one qubit 
at the same state of the computational basis. 
The state $\ket{0011}$ then has
two domains of length two.
A domain wall
is a transition between two domains. 
We do not count transitions
associated to magnons
as domain walls because domains have
more than one qubit.
We illustrate the identification of domain walls and magnons in a state in
Fig.~\ref{fig:dwmagnon}. 
\begin{figure}[ht!]
  \centering
  \begin{tikzpicture}
    \node at (-1,0) {$|$};
    \node at (-0.5,0) {0};
    \node at (0,0) {{\color{blue}1}};
    \node at (0.5,0) {0};
    \node at (1,0) {0};
    \draw [thick] (1.25,-0.3) -- (1.25,0.3);
    \node at (1.5,0) {1};
       \node at (2,0) {1};
    \node at (2.5,0) {1};
    \node at (3,0) {{{\color{red}0}}};
    \node at (3.5,0) {1};
    \node at (4,0) {1};
    \node at (4.5,0) {{\color{red}0}};
   \node at (5,0) {1};
   \node at (5.5,0) {1};
     \draw [thick] (5.75,-0.3) -- (5.75,0.3);
       \node at (6,0) {0};
    \node at (6.5,0) {0};
    \node at (7,0) {{\color{blue}1}};
    \node at (7.5,0) {0};   
    \node at (8,0) {$\rangle$}; 
  \end{tikzpicture}
  \caption{Example of identification of particles, holes, and domain walls. Particles appear in blue, holes in red, and straight vertical lines denote domain walls.}
  \label{fig:dwmagnon}
\end{figure}

Much like the location of magnons themselves,
the positions of domain walls is ill-defined
if there are magnons in their immediate vicinity.
For example, the state
    $\ket{001011}$
has one magnon and one domain wall, and
we can 
say that the particle 
is at the third position
and the domain wall between the fourth and the fifth position,
or a hole at the fourth position
and a domain wall between the second and the third position.
We resolve the ambiguity by choosing magnons to be at the leftmost available position again.
The
state $\ket{001011}$ thus 
has a magnon at the third position 
and domain wall between the fourth and the fifth positions.

Magnons and domain walls allows us to discuss
the basic physical dynamical processes of the model.
The reason is that~\eqref{trans}
can be seen as hopping terms for magnons.
All magnons are separated by at least one
position, and the kinetic term ensures
that this constraint is kept during time evolution.
Neither two
particles nor two holes can jump to neighboring positions.

The states that do not have any domain walls from the
``purely magnonic sector'' of the Hilbert space. 
This sector splits into further subspaces according to the value of
$Q_1$. The purely magnonic sector is the eigenspace with
maximum $Q_2$ for given $Q_1$.
Let $N$ denote the number of qubits (assumed to be even).
The eigenspace with $Q_1=N/2$ has the maximal number $N/2$ of magnons.
No state with domain walls belongs to the $Q_1=N/2$ sector, which
has just two states, namely, the N\'eel state and the
anti-N\'eel state.
Both states are trivial eigenstates of~\eqref{foldedH}
with zero energy.
Purely magnonic sectors with 
$Q_1<N/2$ and $Q_1>N/2$ have non-maximal number of magnons.
The eigenstates in this case
are common to the constrained XX model,
the parity-breaking variant of the folded XXZ model in purely magnonic sectors.
This spin chain is the point
$\Delta=0$ 
of the constrained XXZ 
model~\cite{constrained1,constrained2,constrained3,constrained4},
whose correlation functions have been the object to intensive research~\cite{pronko-abarenkova-ising-limit, bog-ising-vicious,bog-ising-limit,bog-ising-limit2}.
We write the eigenstates of purely magnonic sectors in Subsection~\ref{magnonic}.

The crucial difference between the folded XXZ model
and the constrained XX model, 
in addition to parity invariance, 
is that the former
also allows domain
walls in the Hilbert space.
It follows from the kinetically constrained hopping~\eqref{trans}
that 
isolated domain walls are frozen. For instance, the 
Hamiltonian annihilates the state $\ket{0011}$, 
which has an isolated domain wall
in the middle.
A domain wall only moves when it scatters 
with a magnon, resulting in the non-trivial dynamics
of the model. Let us consider the example of the initial state
$\ket{\psi}=\ket{001001111}$ 
and let us act with the Hamiltonian on top multiple times.
At each step multiple
hopping processes occur. One possible hopping sequence is 
\begin{equation}
  \begin{split}
   \ket{001001111}\quad\rightarrow\quad 
   \ket{000101111}\quad\rightarrow\quad
   \ket{000110111}\quad\rightarrow\quad
 \ket{000111011}\ .
  \end{split}
\end{equation}
We began with a state with a particle on the left and a single isolated and immobile domain wall. The hopping processes dragged the particle close to the domain wall. Since the particle could not reach the domain wall, it ``transmuted'' into a hole instead. (See~\cite{quantum-bowling} for an earlier analysis of such a phenomenon.)
The transmutation of the particle into a hole moved the domain wall to the left by two positions.  
Such interactions happen whenever a magnon meets a domain wall, be it a particle or a hole. 
The scattering event leads to the displacement
of the domain walls as well as the magnons. 
All the eigenstates can be
understood by the interaction of magnons
among them and with the domain walls that might be present.

\subsection{Hilbert Space Fragmentation}

\label{sec:fragm}

The dynamical processes above lead to the phenomenon of Hilbert space fragmentation.
Hilbert space fragmentation denotes the split of the 
Hilbert space splits into exponentially many subspaces,
called fragments or sectors,
such that all the entries of the Hamiltonian between different fragments vanish~\cite{fragmentation-scars-review-2,hfrag-review}. 
The definition is physically meaningful
if the fragments can be constructed
via local constraints or local rules
before actually finding all the exact eigenstates. 

In the folded XXZ model, the fragmentation arises
from the kinematic constraint in the hopping processes. 
Different fragments can be labeled by the numbers of magnons and the number and relative positions of the domain walls. 
The domain walls are not mobile on their own, thus their relative position does not change unless there is a scattering event with a magnon. This implies that, if one properly takes into account the displacements caused by the magnons, then one can uniquely label the fragments by the relative position of the domain walls. 

We should mention that this description only applies to open boundary conditions. 
The mechanism of fragmentation is independent of the boundary conditions, but open boundary conditions enable simple labeling of the fragments.
Such a labeling was introduced in~\cite{Pozsgay21,sajat-hardrod} using ``effective coordinates'' (alternative descriptions of fragmentation were provided earlier in~\cite{folded-XXZ-stochastic-1}). To label the fragments, we follow the effective coordinates of~\cite{Pozsgay21,sajat-hardrod} closely.

Let us then set open boundary conditions, and let us choose a state $\ket{\psi}$ of the computational basis. 
The procedure to obtain the labeling is the following.
First, we compute the action of the Hamiltonian on the state multiple times $H^n\ket{\psi}$.
The result is the Krylov basis of the fragment of $\ket{\psi}$.
If we expand $H^n\ket{\psi}$ into the computational basis, the number of terms of the Krylov basis typically increases with $n$. Our goal is to 
identify a distinguished state of the computational basis appearing at a large enough $n$ that can be used as a label of the fragment. To this aim, we consider all the possible hopping processes which arise from multiple actions of the Hamiltonian, and we focus on a selected direction of the hopping. We move in this way the magnons to their leftmost possible position inside $\ket{\psi}$.
The result is the distinguished state we use to label the fragment, where magnons are tightly packed at the left boundary,
occupying next-to-nearest neighboring positions, while the domain walls remain somewhere else in the spin chain. 

We illustrate the procedure with an example. We consider the initial state
\begin{equation}
   \ket{0001001110111},
\end{equation}
which includes two magnons (one particle and one hole), and one domain wall in the middle.
First, we move the particle on the left side to its leftmost position by applying two hopping processes and obtain
\begin{equation}
   \ket{0100001110111} \ .
\end{equation}
We now move the hole on the right side to the left, transmuting in doing so the hole into a particle and shifting the domain wall in the middle. The result is the state
\begin{equation}
     \ket{\Psi_{\mathrm{label}}}=\ket{0101000011111} \ ,
\end{equation}
which we use as the label of the given fragment. 
Magnons are packed at the left boundary in this state, 
while a single domain wall is left in the chain. 

Given the number of magnons of a sector, we can use the number and the positions of the domain walls
in the distinguished state as a label of the fragment. 
Indeed, 
it follows from the procedure that the final positions of the magnons and the domain wall are uniquely defined. 
There are exponentially many possibilities for the labels for a given the number magnons and domain walls
as per the freedom in placing the domain walls.
This multiplicity that lies at the core of the Hilbert space fragmentation into exponentially many sectors.
Note that magnons in the state that labels fragments are particles if $Q_1<N/2$ and holes if $Q_1>N/2$.
(If $N$ is even, the pair of eigenspaces with $Q_1=N/2$ are one-dimensional and the labeling is trivial
as we have already mentioned.)

It is important to emphasize that the mechanism of fragmentation is not related to the integrability of the model.
For example, the addition of any sort of non-integrable diagonal perturbation would preserve the kinetic constraint of the hopping and thus fragmentation.
Such perturbations were in fact addressed in~\cite{sajat-mpo,maurizio-nonint-jamming}.
Furthermore,~\cite{sajat-mpo} showed that there is a large
non-Abelian symmetry algebra associated with this special type of Hilbert space fragmentation.
The symmetry algebra is generated by matrix-product operators  with fixed bond dimension~\cite{sajat-mpo}, and is compatible even with selected non-integrable perturbations.
We do not use these properties here.

Hilbert space fragmentation has been realized in cold atom experiments~\cite{Scherg_2021, kohlert2021experimentalrealizationfragmentedmodels} 
and engineered in superconducting quantum processors~\cite{Wang24}. 
The implementation of the folded XXZ model using Rydberg atom arrays has been proposed and studied in~\cite{rydberg-folded-exp1, rydberg-folded-exp2}. 
Moreover, Floquet quantum circuits exhibiting Hilbert space fragmentation have been introduced in~\cite{Langlett21, Moudgalya21}.

\section{Eigenstates}

\label{sec:eigenstates}

In this section, we present the spectrum of the folded XXZ model on a chain with open boundaries.
We focus on open boundary conditions because the behavior of domain walls in this case is simpler than for periodic chains.
The labeling of fragments is in fact reflective of the simplicity and will be crucial for the construction of the quantum circuit.  
We consider an open spin chain composed of $N+2$ qubits labeled by $j=0,\ldots,N+1$.
The Hamiltonian
\begin{equation}
  H=-\frac{1}{8}\sum_{j=0}^{N-2}(1+\sigma^z_{j}\sigma^z_{j+3})(\sigma^x_{j+1}\sigma^x_{j+2}+\sigma^y_{j+1}\sigma^y_{j+2}) \ ,
  \label{oH}
\end{equation}
acts diagonally on the boundary qubits at positions $j=0$ and $j=N+1$, as it is clear from the allowed dynamical processes~\eqref{trans}. 
Therefore, boundary qubits carry well-defined quantum numbers
and the Hilbert space splits into four sectors depending on the states of these boundary qubits. The global bit-flip symmetry 
of the Hamiltonian implies that we can fix the qubit at position $j=0$ to be in $|0\rangle$. 

As we explained in Subsection~\ref{sec:fragm}, the Hilbert space fragments of an open chain can be labeled by the state of the computational basis 
where magnons are at their leftmost positions and domain walls lie to their right.
If the last qubit of the chain 
is in $|1\rangle$,
the reference state of a fragment in this sector is
\begin{equation}
|0\,i_1 \dots i_{N-1} 1 1\rangle \hspace{1cm} {\rm or} \hspace{1cm}  |0\,i_1 \dots i_{N-2} 00 1\rangle \ .
\label{refst}
\end{equation}
By using the transition rules~\eqref{trans} again, 
it is simple to see that the fragment associated to the first state is isomorphic to the fragment of a chain with $N+3$ qubits and reference state
\begin{equation}
    |0\,i_1 \dots i_{N-1} 1 10\rangle \ .
\end{equation}
The same argument implies that the fragment labeled by the second state 
is isomorphic to that of a chain with $N+1$ qubits and reference state
\begin{equation}
    |0\,i_1 \dots i_{N-2} 00\rangle \ .
\end{equation}
Hence, without loss of generality, we just consider below the sector where both boundary qubits are in $|0\rangle$.

We consider in the following that states of the form 
\begin{equation}
    |0\, i_1\dots i_{N-2}110\rangle \ , \hspace{1cm} |011 \, i_3\dots i_{N}0\rangle \ ,
\end{equation}
contain a domain wall between the next to last and last qubit, and the first and second qubit respectively,
even though the definition of domain wall of Subsection \ref{sec:dynamics} does not encompass these cases. This guarantees that the number of domain walls in all the states of a fragment is the same. Moreover, it helps to establish a map between the eigenstates of the free XX model and those of the folded XXZ chain which will be the basis for our characterization of the latter. 

With the previous small change in nomenclature,  
we describe now folded XXZ eigenstates of the form $|0\rangle \otimes |\Psi_{M,D}\rangle \otimes |0\rangle$, which carries $M$ magnons and $D$ domain walls.
We will show below that the state $|\Psi_{M,D}\rangle$, describing the central $N$ spins, can be written as
\begin{equation}
    |\Psi_{M,D}\rangle = U_{D} \Big(\, |\Phi_{M} \rangle_{N_0} \otimes |{ 0}\rangle^{\otimes N-N_0} \Big)  \ ,
    \label{wavefunct}
\end{equation}
where $|\Phi_{M} \rangle$ is
an $M$-magnon eigenstate of the XX model, realizing free spinless fermions, and $U_D$ is a unitary. The XX eigenstate lives on an open chain with a reduced number of sites $N_0=N+1-M-D$.
Due to the reflection with the open boundaries,
each magnon of the XX model has components with momenta $\pm p_a$, where
\begin{equation}
    \label{qcond}
    p_a= { \frac{\pi m_a}{N_0+1}} \ , \hspace{1cm} 
\end{equation}
and $m_a$, $a=1,\ldots,M$, are integers.
The unitary operator~$U_D$ introduces the contact repulsion among magnons and contains the information on the domain wall positions. It implements a collection of permutations and bit flips, but does not alter the value of the XX eigenstate coefficients. Therefore, the magnon momenta of the folded XXZ chain are directly inherited from those of $|\Phi_{M} \rangle$.
The only effect of the interaction on the spectrum of allowed momenta is through $M$ and $D$ in the effective length $N_0$ in \eqref{qcond}.

The formula~\eqref{wavefunct} was not given in this form before. It is a convenient representation of the eigenstates because it directly leads to the circuits that prepare them: we prepare $|\Phi_M \rangle$ first, and afterward act with the circuit $U_D$ for which we provide the factorization into elementary gates.
However, 
the key results behind 
formula~\eqref{wavefunct} 
appear in earlier works, including the original work on the constrained XXZ model~\cite{constrained1,constrained2}, 
as well as in the more 
recent~\cite{Pozsgay21,sajat-hardrod}, which treated the folded XXZ model as a ``hard-rod deformation'' of the XX spin chain. 

\subsection{Magnons}

\label{magnonic}

We briefly review purely magnonic eigenstates, 
which are common to the constrained XX model.
Since we have chosen the boundary spins to be in $|0\rangle$, magnons correspond univocally to qubits in $|1\rangle$. 
For states containing a single magnon,
the folded Hamiltonian on a chain of $N+2$ qubits reduces then to that of the XX model on the $N$ bulk qubits.
Therefore, one-magnon 
eigenstates are $|\Psi_{1,0}\rangle=|\Phi_1\rangle$,
where $|\Phi_1\rangle$ is a plane wave with free open boundary conditions:
\begin{equation}
    \label{B1M}
    |\Phi_1\rangle =|\Phi(p)\rangle-|\Phi(-p)\rangle \ , \hspace{1cm} |\Phi(p)\rangle=\sum_{n=1}^N \e^{\im p n} 
    \sigma_n^- |{ 0}\rangle^{\otimes N} \ .
\end{equation}
In this simple case, the unitary $U_0$ in~\eqref{wavefunct} is just the identity.

The folded XXZ Hamiltonian~\eqref{oH} induces free open boundary conditions on the bulk qubits also for states with a general number of magnons. Generalizing \eqref{B1M}, this implies that $|\Psi_{M,0} \rangle$ will be given by 
\begin{equation}
    |\Psi_{M,0} \rangle =\sum_{\epsilon_a =\pm 1}
    \left[\,\prod_{b=1}^M \epsilon_b \right]
    |\Psi(\epsilon_1 p_1,\ldots,\epsilon_M p_M) \rangle  \ .
    \label{XXZmagnons}
\end{equation}
Each $|\Psi(p_1,\ldots,p_M)\rangle$ is a Bethe state, namely,
a superposition of plane waves whose coefficients are determined by the two-body scattering matrix,
which for the folded XXZ model is 
\begin{equation}
  S(p_1,p_2)=-\e^{-\im\,(p_1-p_2)} \ .
\end{equation}
Although this scattering phase represents a non-trivial interaction,
its effect amounts to shifting the coordinate dependence of the plane-wave factors such that
\begin{equation}
  |\Psi(p_1,\ldots,p_M)\rangle=\!\!\!\sum_{1\leq n_1 < \ldots<n_M\leq N} \hspace{-5mm} \Phi(n_1,n_2\!-\!1,\ldots,n_M\!-\!M\!+\!1) \ 
|n_1n_2\ldots n_M\rangle \ ,
\label{Bst}
\end{equation}
where
$\Phi(n_1,\ldots,n_M)$ is the Slater determinant describing free-fermion wavefunctions
\begin{equation}
    \Phi(n_1,n_2,\ldots,n_M) =
    \begin{vmatrix}
        \e^{\im p_1 n_1} & \e^{\im p_1 n_2} &
        \ldots &\e^{\im p_1 n_M} \\
        \vdots & \vdots  & \ddots & \vdots \\
        \e^{\im p_M n_1} &\e^{\im p_M n_2} &
        \ldots &\e^{\im p_M n_M}\\
    \end{vmatrix} \ .
    \label{Slater}
\end{equation}
We stress that~\eqref{Slater} is nothing but the Bethe wavefunction of the CBA for the XX model. In \eqref{Bst} we have introduced the notation
\begin{equation}
|n_1n_2\ldots n_M\rangle=\, \sigma_{n_1}^-\sigma_{n_2}^-\ldots\sigma_{n_M}^-|0\rangle^{\otimes N} \  .
\end{equation}

It is now immediate to define a 
unitary operation embedding
the eigenstates of the 
XX model 
into the purely magnonic sector of the folded XXZ chain. According to~\eqref{wavefunct}, we search for a transformation $U_0$ such that
\begin{equation}
    |\Psi(p_1,\ldots,p_M)\rangle= U_0 \Big( |\Phi(p_1,\ldots,p_M)\rangle_{N_0} \otimes |{ 0}\rangle^{\otimes N-N_0} \Big) \ ,
    \label{U0}
\end{equation}
where in the absence of domain walls $N_0=N-M+1$. From~\eqref{Bst},  
$U_0$ should just shift the position of the free magnons by
\begin{equation}
    \label{shift}
    n_a \mapsto n_a+a-1 \ .
\end{equation}
This mapping shows that the particles of the folded XXZ model behave as ``hard
rods''. In other words, particles, instead of occupying just one position, appear to have a finite length $\ell=2$.
The shift in the coordinates
takes into account the ``accumulation'' of the total length of the particles as counted from the left end of the
spin chain. This coordinate transformation is well-known for the classical hard-rod gas; see, for example,~\cite{hardrod-gge-1}. In the case of the quantum spin chains, this interpretation was given, for instance, in~\cite{Pozsgay21,sajat-hardrod}. 

Substituting~\eqref{U0} into~\eqref{XXZmagnons}, we recover the relation~\eqref{wavefunct} for purely magnonic states. Consistently,
the Bethe equations for the folded XXZ model on a chain with $N+2$ sites impose the same quantization condition on the magnon momenta as that of the XX model on $N_0$ sites \eqref{qcond}.
The states of both chains share the same energy
\begin{equation}
    \label{EM}
    E=-2\sum_{a=1}^{M}\cos p_a \ .
\end{equation}

\subsection{One Magnon and Two Domain Walls}

We construct eigenstates containing
one magnon and two domain walls, since they already
illustrate all the features of the interaction
between magnons and domain walls. 
Our starting point will be the reference state labeling the chosen fragment of Hilbert space, $|\Psi_{\mathrm{label}}\rangle$. As explained in Section~\ref{sec:fragm}, the reference state contains the magnon at its leftmost position, which in this case is just the site at the position $j=1$:
\begin{equation}
   |\Psi_{\mathrm{label}}\rangle= \sigma_1^- |\Psi_{0,2} \rangle \ .
    \label{ref12}
\end{equation}
The state $|\Psi_{0,2} \rangle$ describes the domain-wall positions.
Let the left domain wall
lies between
the positions
$d_1$ and $d_1+1$,
and the right domain wall lies
between
the positions $d_2$ and $d_2+1$. Then
\begin{equation}
  |\Psi_{0,2}\rangle =\prod_{j=d_1+1}^{d_2}\sigma_j^- \, |0\rangle ^{\otimes N}\ ,
\end{equation}
where necessarily $2 \leq d_1<d_2\leq N$. 

Starting from~\eqref{ref12}, the magnon behaves as a particle that propagates
freely until position $d_1-1$, that is, until it faces the first domain wall.
We denote the basis of available states to the left of the domain wall as
\begin{equation}
    |\psi_2(n)\rangle_{\mathrm{LP}}=\sigma_n^-|\Psi_{0,2}\rangle \ .
\end{equation}
The magnon cannot reach the position $d_1$ as a particle, since 
the 
transition rules~\eqref{trans}
forbid the merger between
a particle 
and a domain. This forces the magnon to transmute into a hole,
a process that displaces the 
first
domain wall two positions to the left, between positions
$d_1-2$ and $d_1-1$.
The magnon propagates then
freely as a hole between $d_1+1\leq n \leq d_2-1$. We represent these configurations by
\begin{equation}
    |\psi_2(n)\rangle_{\mathrm{H}}=\sigma_{d_1-1}^-\sigma_{d_1}^-\sigma_n^+|\Psi_{0,2}\rangle \ .
\end{equation}
The magnon abandons the domain after the position $d_2-1$, turning into a particle again.
This shifts the right
domain wall to the position
between  the $d_2-2$ and $d_2-1$ sites. 
Finally, the magnon propagates
freely between $d_2+1\leq n\leq N$. We represent the corresponding states by
\begin{equation}
    |\psi_2(n)\rangle_{\mathrm{RP}}=\sigma_{d_1-1}^-\sigma_{d_1}^-\sigma_{d_2-1}^+\sigma_{d_2}^+
    \sigma_n^-|\Psi_{0,2}\rangle \ .
\end{equation}
The magnon in summary propagates as 
a free spinless fermion,
whose position receives additional shifts
each time it scatters with a domain wall, in turn displaced as well.

According to the previous description, the eigenstates of one magnon and two domain walls 
break
into three parts: the left particle,
the hole, and the right particle. Explicitly,
\begin{equation}
    |\Psi_{1,2}\rangle=|\Psi_{\mathrm{LP}}\rangle+
    |\Psi_{\mathrm{H}}\rangle+|\Psi_{\mathrm{RP}}\rangle \ .
\end{equation}
Each term separately mimics the
state~\eqref{B1M}
over a sub-chain:
\begin{equation}
    \begin{split}
    &|\Psi_{\mathrm{LP}}\rangle=
    \sum_{n=1}^{d_1-1} \sum_{\epsilon=\pm1}
    \epsilon \e^{\im\epsilon p n}
    |\psi_2(n)\rangle_{\mathrm{LP}} \ , \\
    &|\Psi_{\mathrm{H}}\rangle=
    \sum_{n=d_1}^{d_2-2}\sum_{\epsilon=\pm1}\epsilon \e^{\im\epsilon p n}
    |\psi_2(n+1)\rangle_{\mathrm{H}}\ , \\
    &|\Psi_{\mathrm{RP}}\rangle=
    \sum_{n=d_2-1}^{N-2}\sum_{\epsilon=\pm1}\epsilon \e^{\im\epsilon p n}
    |\psi_2(n+2)\rangle_{\mathrm{RP}}\ .
\end{split}
\end{equation}
The Bethe equations require the magnon momentum to satisfy
\begin{equation}
    p =\frac{\pi m}{N-1} \ ,
    \label{p12}
    \end{equation}
with $m$ integer, which coincides with momentum quantization of a
free spinless fermion propagating on an open chain of $N_0=N-2$ qubits. Domain walls carry
no momenta since they are static but for scattering with magnons.

It is straightforward to define 
a unitary operator satisfying
\begin{equation}
    \label{Psi12}
    |\Psi_{1,2}\rangle=U_2\Big( |\Phi_1\rangle_{N-2}\otimes |00\rangle \Big)  \ ,
\end{equation}
where $|\Phi_1\rangle$ is the plane wave~\eqref{B1M}. The operator $U_2$ knows about the position of domain walls in the reference state~\eqref{ref12}, and acts as follows
\begin{equation}
        \sigma_n^-|0\rangle^{\otimes N-2} \;\;\; \longrightarrow \;\;
        \begin{cases}
            \; |\psi_2(n)\rangle_{\mathrm{LP}} \ ,  & \hspace{.4cm} 1 \leq n\leq d_1-1 \ ,\\
            \; |\psi_2(n+1)\rangle_{\mathrm{H}} \ , & \hspace{.4cm} d_1 \leq n \leq d_2-2 \ ,\\
            \; |\psi_2(n+2)\rangle_{\mathrm{RP}} \ , & \hspace{.4cm} d_2-1\leq n \leq N-2 \ .
        \end{cases}
        \label{U2}
\end{equation}
The operator $U_2$ detects the position of the magnon on the shorter free chain with 
no domains, relocates the magnon into the physical chain containing two domain walls at their reference positions, 
and shifts the domain walls when required according to the rules~\eqref{trans}. 
We close the analysis by stressing that the energy of the eigenstate~\eqref{Psi12} is again given by~\eqref{EM} with $M=1$.
Domain walls correct the quantization of momenta, but do not affect the form of the dispersion relation.

\subsection[\texorpdfstring{$M$}{M} Magnons and \texorpdfstring{$D$}{D} Domain Walls]{\texorpdfstring{$\boldsymbol{M}$}{M} Magnons and \texorpdfstring{$\boldsymbol{D}$}{D} Domain Walls}

\label{gen}

Finally, we build eigenstates containing
$M$ magnons and $D$ domain walls.
These states
combine the features of the
interaction of magnons among themselves and with
domain walls treated in the previous subsections.

As before, our starting point is the reference state labeling the fragment to which the eigenstate we want to prepare belongs, $|0\rangle \otimes |\Psi_{\mathrm{label}}\rangle \otimes |0\rangle$. This state packs the magnons to the left, taking into account that they cannot occupy neighboring positions, and thus reads
\begin{equation}
|\Psi_{\mathrm{label}}\rangle=\sigma_1^-\sigma_3^-\ldots\sigma_{2M-1}^-|\Psi_{0,D}\rangle \ .
\label{refgen}
\end{equation}
The state $|\Psi_{0,D}\rangle$ carries the domain walls at their rightmost positions.
The number of domain walls is always even because boundary qubits are at $\ket{0}$. Let the first domain wall
lie between
the positions
$d_1$ and $d_1+1$, where necessarily $d_1>2M$, the second
between
$d_2$ and $d_2+1$, and so on. The domain walls reference  state reads
\begin{equation}
|\Psi_{0,D}\rangle=
\left[\prod_{j_1={d_1+1}}^{d_2}\sigma_{j_1}^-\right]\left[\prod_{j_2={d_3+1}}^{d_4}\sigma_{j_2}^-\right]\ldots\left[\prod_{j_{D/2}=d_{D-1}+1}^{d_D}\sigma_{j_{D/2}}^-\right] |0\rangle^{\otimes N} \ .
\label{refD}
\end{equation}

The dynamical connection between the state that labels the fragment and other configurations parallels the simple case addressed in the previous subsection. 
The only new constraint is the contact repulsion among magnons. We illustrate the situation with an example.
Let us assume
the rightmost $L$ magnons are inside the (large enough) domain that lies between position $d_K$ and $d_{K+1}$, with $K$ odd, while
the remaining $M-L$ magnon still lie at their initial positions. 
The associated state is
\begin{equation}
    \underbrace{\sigma_{1}^-\sigma_3^-\ldots\sigma_{2M-2L+1}^-}_{M-L~\text{particles}}\underbrace{\sigma_{d_1-2L}^{\,x}\sigma_{d_1-2L+1}^{\,x}\ldots \sigma_{d_K-2}^{\,x}\sigma_{d_K-1}^{\,x}}_{{K~\text{displaced domain walls}}}\underbrace{\sigma_{h_1}^+\sigma_{h_2}^+\ldots
    \sigma_{h_L}^+}_{L~\text{holes}}|\Psi_{0,D}\rangle \ ,
\end{equation}
where the locations of the holes satisfy the criterion of contact repulsion $h_i < h_{i+1}-1$.
We have used $\sigma^{\,x}_j$ to relocate domain walls since
it correctly acts as $\sigma^-_j$ or $\sigma_j^+$  when shifting the beginning or end of a domain, respectively.
The remaining configurations with particles, holes, and domains walls are analogous.
Particles and holes are placed within domains, and domain walls shift according 
to the number of magnons that have crossed them. 

Like for 
purely magnonic eigenstates, 
the wavefunction of generic eigenstates is 
a Slater determinant whose dependence on magnon positions is properly shifted.
The folded XXZ magnons behave as hard rods with effective length two.
The effective length that the magnons perceive is further shortened by the presence of the domain walls. As illustrated in \eqref{U2},
the transmutation between particles and holes each time a magnon encounters a domain wall 
entails an additional shift of magnons one step to the right of the initial position of the domain wall.
For this reason, it is enough to define the free-fermionic seed eigenstate $|\Phi_M\rangle$ in~\eqref{wavefunct} on a chain of length $N_0=N-M+1-D$.
To obtain the circuit preparing folded XXZ eigenstates,
we find convenient to split the operator $U_D$ into the product of $U_0$, which
constructs a purely magnonic state, and $V_D$, which introduces the domain walls next. 
In other words,
\begin{equation}
\label{VD}
|\Psi_{M,D}\rangle=U_D\Big(|\Phi_M\rangle_{N_0} \otimes |0\rangle^{\otimes N-N_0}\Big)=V_D\Big(|\Psi_{M,0}\rangle_{N-D} \otimes |0\rangle^{\otimes D} \Big) \ .
\end{equation}

Given a component $|n_1 n_2 \dots n_M\rangle$ of the state $|\Psi_{M,0}\rangle$,
the unitary $V_D$ acts in the following iterative way. First, we focus on the rightmost qubit in $|1\rangle$, at position $n_M$.
The action of $V_D$ is the insertion of a magnon in the state with domain walls like
\begin{equation}
  |\psi_{D}(n_M)\rangle= \left[\prod_{a=1}^{k_M}\sigma_{d_a-1}^{\,x}\sigma_{d_a}^{\,x} \right] \sigma^{\,x}_{n_M+k_M} |\Psi_{0,D}\rangle \ ,
  \label{reloc}
\end{equation}
where
\begin{equation}
    \begin{array}{ll}
        k_M=0 \ , & \hspace{4mm} n_M< d_1 \ ,\\
        k_M=a \ , & \hspace{4mm}d_a < n_M+a < d_{a+1} \ , \\
        k_M=D \ , & \hspace{4mm} d_D < n_M+D \ .
    \end{array} 
    \label{displace}
\end{equation}
The particle at the position $n_M$ in the magnonic state moves to position $n_M+k_M$ in the presence of the domain walls,
remaining a particle if $k_M$ is even and transmuting into a hole if $k_M$ is odd.
At the same time, the domain walls shift by $d_a\mapsto d_a-2$ for $a\leq k_M$
and stay at their original positions otherwise. Next, $V_D$ inserts
the particle of the purely magnonic state at position $n_{M-1}$
according to~\eqref{reloc} again, but with $|\Psi_{0,D}\rangle$
replaced by $|\psi_D(n_M)\rangle$. This step generates a new state $|\psi_D(n_{M-1},n_M)\rangle$, in which
the displacement $k_{M-1}$ is determined by~\eqref{displace} with respect to the new positions of the domain walls.
The unitary $V_D$ iterates the process until $|\psi_D(n_1,\dots,n_M)\rangle$ is obtained.
Analogously to~\eqref{Bst}, the eigenstates in the fragment correspond to the superposition
\begin{equation}
    |\Psi_{M,D} \rangle =\sum_{\epsilon_a =\pm 1} 
    \left[\,\prod_{b=1}^M \epsilon_b \right]
    |\Psi_D(\epsilon_1 p_1,\ldots,\epsilon_M p_M) \rangle  \ ,
    \label{XXZmagnonsDW}
\end{equation}
with
\begin{equation}
  |\Psi_D(p_1,\ldots,p_M)\rangle=\!\!\!\sum_{1\leq n_1 < \ldots<n_M\leq N} \hspace{-5mm} \Phi(n_1,n_2\!-\!1,\ldots,n_M\!-\!M\!+\!1) \,
|\psi_D(n_1,n_2,\ldots ,n_M)\rangle \ ,
\label{Bstgen}
\end{equation}
and $\Phi(n_1,\dots,n_M)$ the Slater determinant~\eqref{Slater}.
The energy is given by~\eqref{EM}

\section{The Quantum Algorithm}

\label{squantalg}

In this section, we build quantum circuits that efficiently prepare arbitrary eigenstates of the folded XXZ model. 
Following \eqref{wavefunct} and \eqref{VD}, 
the circuits consist of three parts. The first part builds eigenstates
of the XX model with free open boundaries. 
The second transforms the eigenstates of
the XX model into purely
magnonic states of the folded XXZ chain.
This part realizes the unitary $U_0$ defined in Subsection~\ref{magnonic}.
Finally, the third part implements the unitary~$V_D$ introduced in Subsection~\ref{gen}, which adds domain walls at the
appropriate positions. 

\subsection{Algebraic Bethe Circuits with Open Boundary Conditions}

\label{ssOFFABC}

Several efficient quantum circuits for preparing free-fermion eigenstates have been proposed in the 
literature~\cite{Verstraete09,Jiang17,Kivlichan18,CerveraLierta18,Farreras24}. 
In this paper, we build upon the recent approach introduced in~\cite{Sopena22,Ruiz23}, 
which successfully recasts the Bethe Ansatz into a quantum circuit known as ``Algebraic Bethe Circuit'' (ABC). 
The computational resources required for ABC scale linearly with the number of qubits but, 
in general, 
exponentially with the number of magnons. 
However, 
for the free XX model, 
the scaling is linear in both qubits and magnons, matching the efficiency of previous algorithms~\cite{Jiang17,Kivlichan18}. 
The works of~\cite{Sopena22,Ruiz23} focused specifically on preparing Bethe states for periodic chains. 
In Appendix~\ref{appOABC}, we show that this construction readily generalizes to open boundary conditions. 
Here, 
we present the quantum circuit and refer to the appendix for further details.

\begin{figure}[ht!]
    \begin{subfigure}[b]{0.33\textwidth}
        \centering
        \includegraphics[scale=0.85,angle=90]{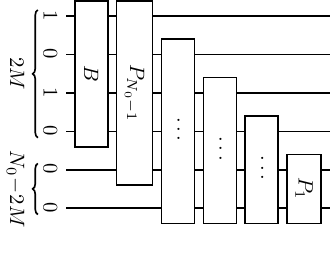}
        \caption{}
    \label{figABCqc}
    \end{subfigure}
    \hspace{-10mm}  
    \begin{subfigure}[b]{0.33\textwidth}
        \centering
        \includegraphics[scale=0.85,angle=90]{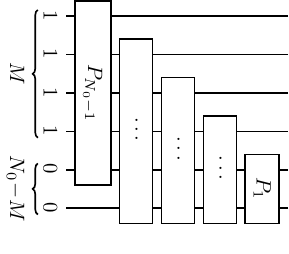}
        \caption{}
    \label{figABCqc2}
    \end{subfigure}
    \hspace{-10mm}  
    \begin{subfigure}[b]{0.33\textwidth}
        \centering
        \includegraphics[scale=0.85,angle=90]{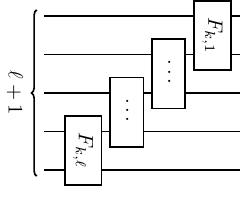}
        \caption{}
    \label{figPF}
    \end{subfigure}
    \centering
    \caption{
    Structure of the Algebraic Bethe Circuits. 
    Fig.~\ref{figABCqc} depicts the quantum circuit for open boundary conditions.
    Fig.~\ref{figABCqc2} depicts the quantum circuit f\textbf{}or periodic boundary conditions.
    Fig.~\ref{figPF} shows the decomposition of the unitaries $P_k$ into two-qubit gates for the XX model, where $\ell=\min(k,2M)$.
    }
    \label{figABC}
\end{figure}

The quantum circuit prepares normalized Bethe states of the XX model with $M$ magnons and free open boundaries over $N_0$ sites, namely
\begin{equation}
    |\widehat{\Phi}_M\rangle
    = \frac{1}{\sqrt{\langle\Phi_M |\Phi_M\rangle}}
    \sum_{\epsilon_a=\pm1}
    \left[\,
    \prod_{b=1}^M \epsilon_b
    \right]|\Phi(\epsilon_1 p_1,\ldots, \epsilon_M p_M ) \rangle \ ,
    \label{superp2}
\end{equation}
where~$|\Phi(p_1,\ldots,p_M)\rangle$ 
denote the Bethe states of the periodic XX model, 
whose wavefunctions are given in~\eqref{Slater}.
The corresponding ABC consists 
of~$N_0$ unitaries arranged in the staircase pattern shown in Fig.~\ref{figABCqc}. The ABC for the periodic boundary conditions is included in Fig.~\ref{figABCqc2} for comparison.
The unitaries~$P_k$ have the same structure in both cases~\cite{Ruiz23}, but those for open boundaries act
 on $2M+1$ instead of $M+1$ qubits. 
This difference is due to the dependence of \eqref{superp2} on both $p_a$ and $-p_a$.
The larger size of the gates however does not 
imply a relevant increase in complexity, since for the XX model each $P_k$
factorizes into a single layer of  two-qubit unitaries as per Fig.~\ref{figPF}. These two-qubit unitaries are of a particular type denoted matchgates, which are characeristic of free-fermionic circuits \cite{Jozsa08} and whose analytical expression
can be found in Appendix~\ref{appMG}. The open ABC contains an additional gate $B$, acting on $2M$ qubits initialized in a reference N{\'e}el state, with
the role of implementing the effect of the open boundaries. A decomposition of $B$ in terms of matchgates is given in Appendix~\ref{appOABC}, with their parameters determined in this case variationaly~\cite{code_xxz_folded}.
The number of gates required by each unitary is summarized in Table~\ref{tabmg}. It is worth mentioning that the ABC can prepare the state \eqref{Slater} with arbitrary momenta, and thus the quantization condition \eqref{qcond} must be imposed separately. 

\begin{table}[ht!]
    \centering
    \renewcommand{\arraystretch}{1.5}
    \setlength{\tabcolsep}{5pt}
    \small
    \begin{tabular}{c|ccc}
    \hline \hline 
     & $P_k$ & $B$\\
     \hline
     Matchgates & $\min(k+1,2M)$ & $(M-2)(2M+1)$
     \\
    \hline
    \end{tabular}
    \caption{Number of matchgates per unitary of the ABC preparing free fermion eigenstates with open boundaries. When $M<2$, the
    decomposition of $B$ requires only one and three matchgates respectively, which is not described by the above formula.}
    \label{tabmg}
    \end{table}

\subsection{From Free to Contact-Repulsive Magnons}

\label{ssFreeContact}

We construct now a quantum circuit that maps XX eigenstates 
into purely magnonic eigenstates of the folded XXZ model, realizing the unitary $U_0$ we detailed in Subsection \ref{magnonic}:
\begin{equation}
     |\Psi_{M,0}\rangle = U_{0} \Big(\, |\Phi_{M} \rangle_{N-M+1} \otimes |{ 0}\rangle^{\otimes M-1} \Big) \ .
\end{equation}
This transformation introduces a contact repulsion among magnons by shifting free-magnon positions according to~\eqref{shift}. Recall that we focus on folded XXZ eigenstates  
with the non-dynamical boundary spins at $|0\rangle$, such that $|\Psi_{M,0}\rangle$ denotes the state of the $N$ bulk spins of the chain. The qubit requirements of the circuit are  the following.
\begin{itemize}
\item A physical 
register~$R_{\mathrm{phys}}$ 
with~$N$ 
qubits initialized in the state $|\Phi_{M} \rangle \otimes |{0}\rangle^{\otimes M-1}$.

\item An auxiliary register $R_{\mathrm{aux}}$ with $M+1$ qubits initialized in $|1\rangle |0\rangle^{\otimes M}$, which keeps track of how many magnons have been already shifted.    
\end{itemize}

\begin{figure}[ht!]
\centering
  \hspace{-10mm}  \begin{subfigure}[t]{0.5\textwidth}
        \centering
        \includegraphics[scale=0.7,angle=90]{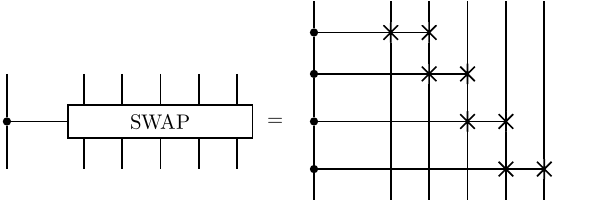}
        \caption{}
        \label{figSWAP}
    \end{subfigure}
  \hspace{-7mm}  \begin{subfigure}[t]{0.5\textwidth}
        \centering
        \includegraphics[scale=0.7,angle=90]{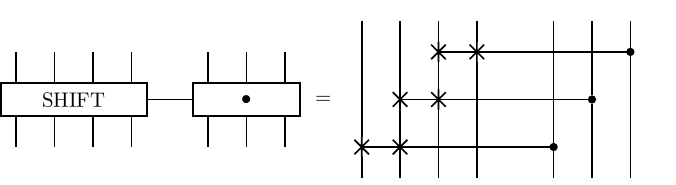}
        \caption{}
        \label{figSHIFT}
    \end{subfigure}
    \caption{Building blocks of the circuit that implements the unitary $U_0$.}
    \label{figU0mod}
\end{figure}

The circuit searches for the free magnons in $|\Phi_M\rangle$ scanning from right to left.
Every time a magnon is found, the qubit in $|1\rangle$ in $R_{\mathrm{aux}}$ moves one position toward the right. The concatenation of $\mathrm{CSWAP}$ gates of Fig.~\ref{figSWAP} performs this operation. 
We describe the action of the circuit on states of the computational basis. Consider that all operations associated to the scan from positions $N-M+1$ down to $n+1$ of $R_{\mathrm{phys}}$ have been already performed. The circuit focuses next on the $n$-th qubit, where we assume that the $r$-th magnon is placed. The module in Fig.~\ref{figSWAP} is executed with the $n$-th qubit as control. This places the $|1\rangle$ in $R_{\mathrm{aux}}$ at position $M-r+2$.  
The circuit now use the information in the auxiliary register to shift the $r$-the magnon according to~\eqref{shift}, which is achieved by applying the module of Fig.~\ref{figSHIFT} to the qubits $n$ to $n+M-1$ of $R_{\mathrm{phys}}$. Note that the action of this module is trivial if there is no magnon at position $n$. 

\begin{figure}[ht!]
\centering
\includegraphics[scale=0.75,angle=90]{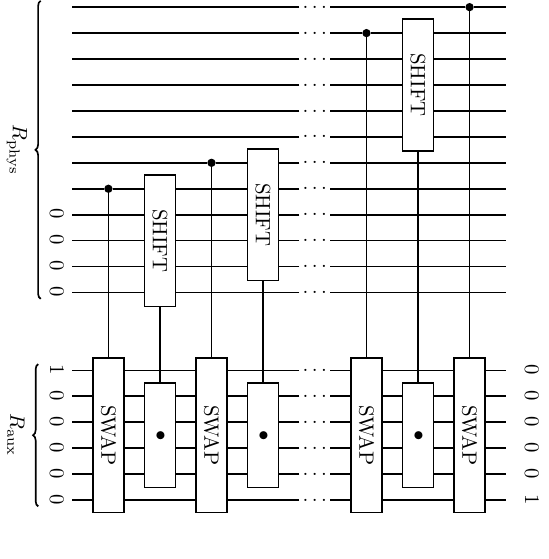}
    \caption{Circuit of the unitary $U_0$. It comprises fewer than $(2M\!-\!1)(N\!-\!M+1)$ $\mathrm{CSWAP}$ gates.
    For simplicity, we assume that all $\mathrm{CSWAP}$ gates act on $M+1$ qubits, which results in an overestimate of the total number of $\mathrm{CSWAP}$ gates.}
    \label{figU0qc}
\end{figure}

The iterating these operations leads us to the quantum circuit depicted in Fig.~\ref{figU0qc}. Any $M$-magnon initial state leaves $R_{\mathrm{aux}}$ in the state $ |0\rangle^{\otimes M}|1\rangle$ at the end of the circuit. This guarantees that physical and ancillary qubits end in a product state.
Let us mention that the circuit in Fig.~\ref{figU0qc} admits straightforward simplifications that we have not included for the sake of clarity, since the modules in Figs.~\ref{figSWAP}--\ref{figSHIFT}
at the beginning and end of the circuit can only have a non-trivial effect on a reduced number of qubits.

\subsection{Adding Domain Walls}

\label{addingDW}

A folded XXZ eigenstate with $D$ domain walls can be obtained by applying the
unitary $V_D$ to the purely magnonic state $|\Psi_{M,0}\rangle$: 
\begin{equation}
\label{VDqc}
|\Psi_{M,D}\rangle=V_D\Big(|\Psi_{M,0}\rangle_{N-D} \otimes |0\rangle^{\otimes D} \Big) \ .
\end{equation}
The unitary $V_D$ is defined by the shifting rules defined by~\eqref{displace}. 
We explain a circuit implementation of $V_D$ that requires 
 $2N+5$ qubits. The qubits are grouped as follows.

\begin{itemize}

\item A physical register $R_{\mathrm{phys}}$ with $N$ qubits initialized in the state $|\Psi_{0,D}\rangle$, which carries the domain walls at their positions in the reference state~\eqref{refD}. 

\item An auxiliary register $R_{\mathrm{aux}}$ containing $N-D$ qubits initialized in the state $|\Psi_{M,0}\rangle$, which  corresponds to the physical register of the previous subsection.

\item A two-qubit register $R_{\mathrm{0}}$ initialized in $|0\rangle^{\otimes 2}$ used as workspace.

\item A register $R_{\mathrm{c}}$ with $D/2+2$ qubits initialized in $|01\rangle |0\rangle^{\otimes {D/2}}$, which counts how many domain walls lie to the left of a given position in $R_{\mathrm{phys}}$.

\item An additional $D/2+1$ register $R_{r}$
initialized in $|1\rangle |0\rangle^{\otimes {D/2}}$, which guarantees that $R_{\mathrm{phys}}$ is in a product state with any other registers of the circuit output.

\end{itemize}

The circuit deploys the magnons in $R_{\mathrm{aux}}$ successively, scanning from right to left, and inserts them into
$R_{\mathrm{phys}}$, shifting along the way the domain walls according to the rules~\eqref{displace}. We describe the action of the circuit when $R_{\mathrm{aux}}$ is initialized in a computational basis state compatible with the contact repulsion among magnons.
Let us consider that all operations associated to the scan from the rightmost end to the $n+1$-th qubit of $R_{\mathrm{aux}}$ have been performed, and denote by $d_1,\dots, d_D$ the positions of the domain walls in $R_{\mathrm{phys}}$ at this point. The focus turns then to
the $n$-th position of the auxiliary register, $R_{\mathrm{aux},n}$. The tasks to be performed can be divided in three steps: 
moving the domain walls if a magnon is found at $R_{\mathrm{aux},n}$, inserting the magnon into $R_{\mathrm{phys}}$, and resetting the auxiliary qubits.

\paragraph{Moving the domain walls.}

\begin{figure}[ht!]
\centering
\includegraphics[scale=0.75,angle=90]{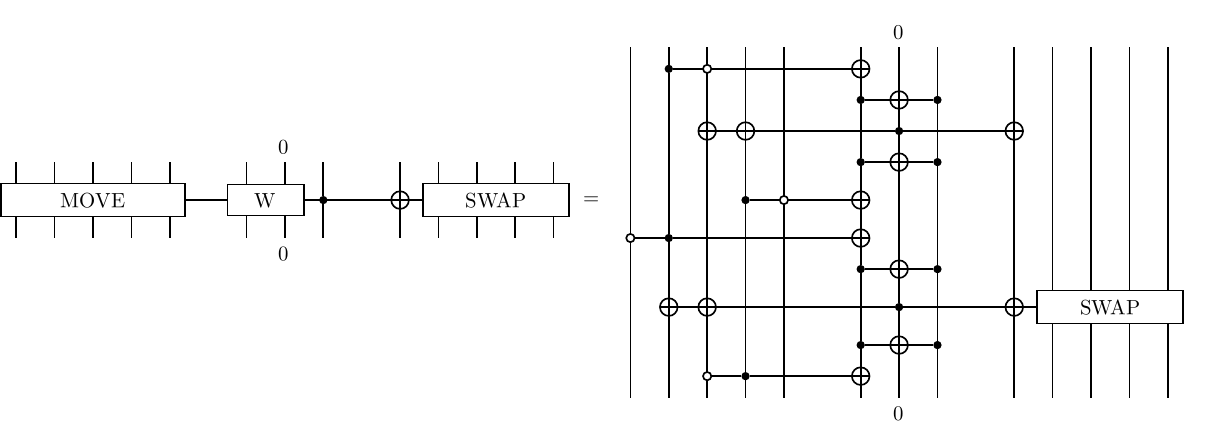}
   \caption{\label{figmove}Module that shifts domain walls of the type  $\ket{01}$ and $\ket{10}$.}
\end{figure}

The basic element of the circuit that relocates the domain walls is  the module shown in Fig.~\ref{figmove}, which searches in the first block of qubits for $\ket{01}$ and  $\ket{10}$
domain walls at positions 3 and 4 respectively, and moves them two sites toward the right depending on the state of a control qubit. If the control qubit is active, the 9-th qubit flips every time a domain wall is found, while the last $D/2+1$ qubits count the number of $|01\rangle$ domain walls.
Recall that the presence of a magnon at $R_{\mathrm{aux},n}$ does not automatically trigger a shift of domain walls. Only those at positions $d_1,\ldots,d_k$ must be shifted, where $k$ satisfies \eqref{displace}, that is,
\begin{equation}
    d_k<n+k<d_{k+1} \ .
    \label{k}
\end{equation}
The control qubit should take care both of this condition and the state at $R_{\mathrm{aux},n}$.

\begin{figure}[ht!]
\centering
\includegraphics[scale=0.75,angle=90]{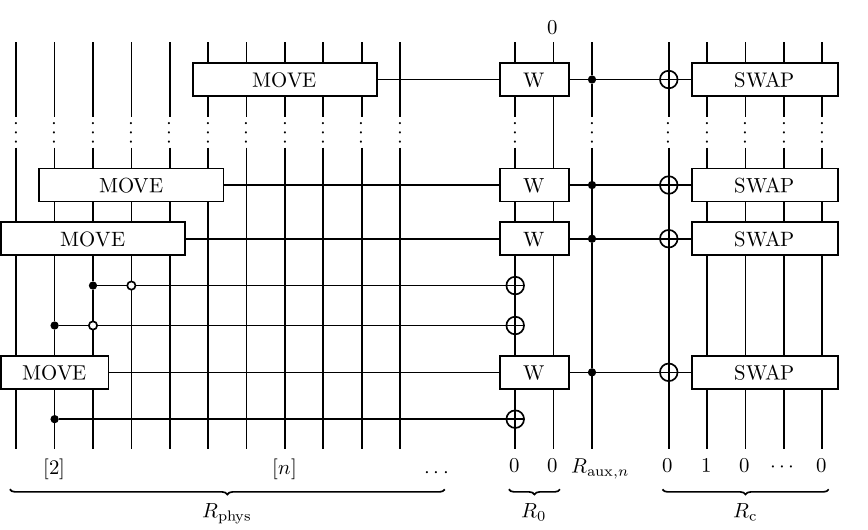}
    \caption{\label{figDWmod1} Circuit that shifts and counts domain walls starting from the first qubit in $R_{\mathrm{phys}}$ up to positions $n$ ($|01\rangle$) and $n + 1$ ($|10\rangle$).
    }
\end{figure}
\begin{figure}[ht!]
\centering
\includegraphics[scale=0.75,angle=90]{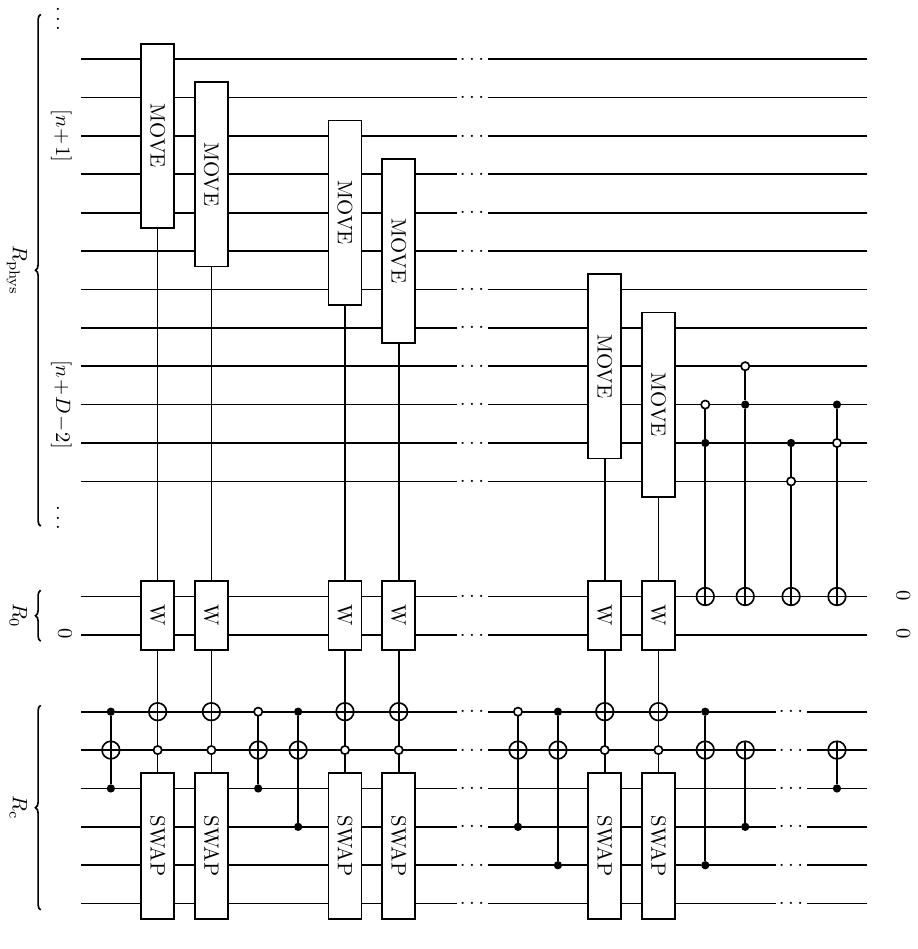}
    \caption{\label{figDWmod2}Circuit that 
    completes the shift and count of domain walls by exploring up to $n+D$, the rightmost possible final position of the magnon.
    }
\end{figure}

The relocation of domain walls proceeds in two steps.
In the first step, shown in Fig.~\ref{figDWmod1}, the  circuit searches for $\ket{01}$ and  $\ket{10}$
domain walls from the first qubit in $R_{\mathrm{phys}}$ up to positions $n$ and $n+1$  respectively. 
The domain walls encountered in this scan must be displaced if a magnon sits at $R_{\mathrm{aux},n}$, which hence is used as control for the gates defined in Fig.~\ref{figmove}. The register $R_{\mathrm{c}}$ stores the information on the number of domain walls found, whose total number we denote by $k'$. Condition \eqref{k} implies that $k'=k$ only if $k' \leq 1$. Otherwise the search for domain walls must be extended at least until the position $n+k'$. If no further domain wall is found, \eqref{k} ensures that $k'=k$ and the search stops. Alternatively the circuit updates $k'$ to be the total number of domain walls up to that point, and iterates the previous operations. The circuit in Fig.~\ref{figDWmod2} implements this scheme. 
Contrary to that in 
Fig.~\ref{figDWmod1}, 
it uses the second qubit of $R_{\mathrm{c}}$ as control for the shifting and counting of domain walls. This qubit is considered active at $|0\rangle$ instead of $|1\rangle$, which  automatically ensures that there is a magnon at $R_{\mathrm{aux},n}$ and $k'\geq 1$. It should be flipped from $|0\rangle$ to $|1\rangle$ when the $k$-th domain wall is reached. This is achieved by a series of Toffoli gates acting on the appropriate qubits of $R_{\mathrm{c}}$. At the end of the circuit, the second qubit of $R_{\mathrm{c}}$ is reset to  
its input state by a last Toffoli gate and a series of $\mathrm{CNOT}$s.

The operations performed by the module in Fig.~\ref{figmove} are carried out with the help of an ancillary workspace comprised of two qubits. This is symbolically represented by the gate W, a notation which we will keep using along this section. 
The working qubits are provided by the register $R_{\mathrm{0}}$.
The second qubit of $R_{\mathrm{0}}$ resets to $|0\rangle$ after completion of the module in Fig.~\ref{figmove}.
Although it does not trigger any undesired operation,
the first qubit does not reset in general when the control qubit is not active. This problem could be solved by adding two more Toffoli gates per module. A more resource-efficient solution is to have the first qubit return to $|0\rangle$ after the circuits of Figs.~\ref{figDWmod1}  and \ref{figDWmod2} have been completed. This is ensured by the $\mathrm{CNOT}$ and Toffoli gates at the beginning of Fig.~\ref{figDWmod1} and  the four Toffoli gates at the end of Fig.~\ref{figDWmod2}. These additional gates only play a role when the control is not active, forcing the concatenated MOVE gates to find every domain wall twice and thus flipping an even number of times the first qubit of $R_{\mathrm{0}}$.

Let us conclude with a remark on the circuit depicted in Fig.~\ref{figDWmod1}. Its first MOVE gate is centered in the second qubit of $R_{\mathrm{phys}}$. According to the pattern in Fig.~\ref{figmove}, the leftmost qubit of this gate coincides with the left boundary qubit of the spin chain, which, being frozen at $|0\rangle$ by construction, has been omitted from $R_{\mathrm{phys}}$. The presence of a magnon at $R_{\mathrm{aux},n}$ forces any $\ket{10}$
domain wall to lie at $d\geq 4$, out of reach of the first MOVE gate, rendering thus the operations associated to the search for this type of domain wall irrelevant. As a result of these simplifications, the range of the first MOVE gate can be reduced from five to three qubits.

\paragraph{Inserting the magnon into $R_{\mathrm{phys}}$.}

\begin{figure}[hb!]
\centering
\includegraphics[scale=0.75,angle=90]{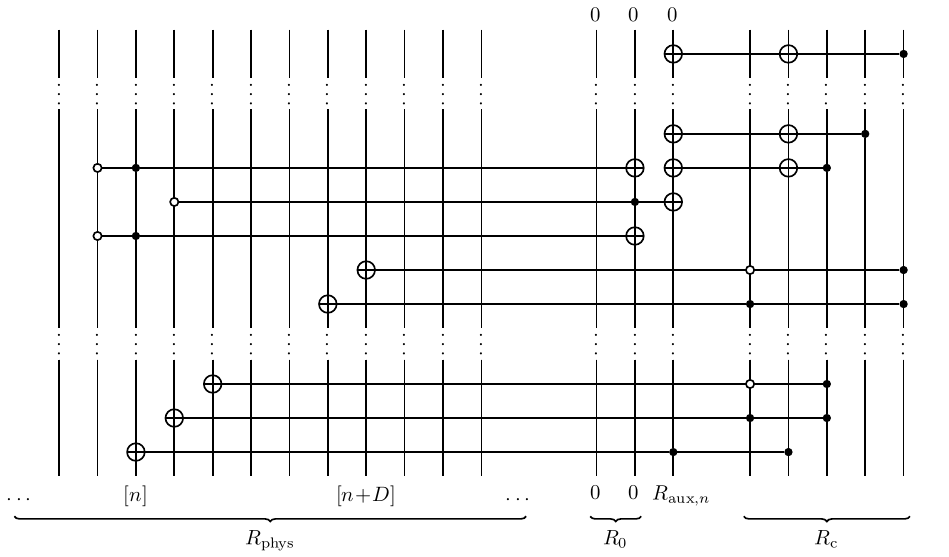}
    \caption{\label{figinsert}Circuit adding the magnons into $R_{\mathrm{phys}}$.}
\end{figure}

The circuits of Figs.~\ref{figDWmod1} and \ref{figDWmod2} correctly relocate the domain walls according to the potential presence of a magnon in $R_{\mathrm{aux},n}$. The next step is to move the magnon, if present, 
from $R_{\mathrm{aux},n}$ into $R_{\mathrm{phys}}$ with the help of the information stored in $R_{\mathrm{c}}$. The circuit in Fig.~\ref{figinsert} fulfills this task.
The first chain of Toffoli gates duplicates the magnon and places the duplicate at the appropriate location in $R_{\mathrm{phys}}$. A further set of Toffoli gates resets $R_{\mathrm{aux},n}$ to $|0\rangle$.

\paragraph{Reset of ancillary qubits.}

\begin{figure}[hb!]
\centering
    \begin{subfigure}[hb!]{\textwidth}
        \centering
        \includegraphics[scale=0.75,angle=90]{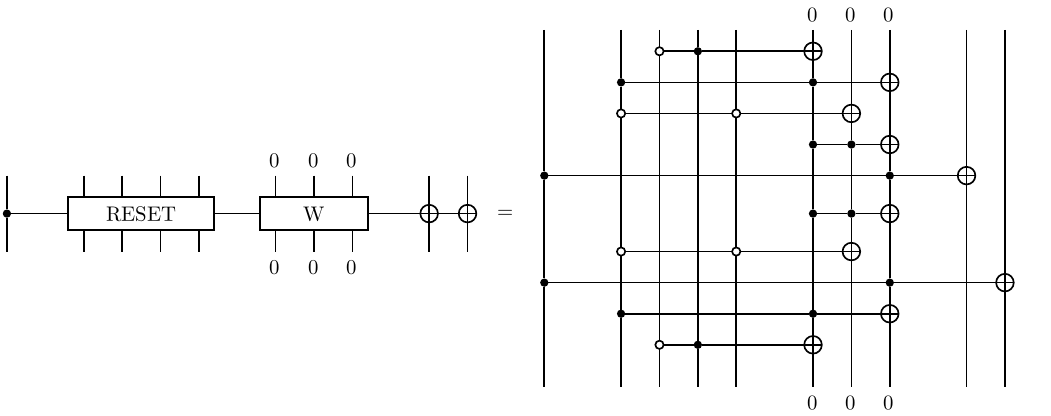}
        \caption{}
        \label{figreset_w}
    \end{subfigure}
    \vspace{5pt}
    \begin{subfigure}[hb!]{\textwidth}
        \centering
        \includegraphics[scale=0.75,angle=90]{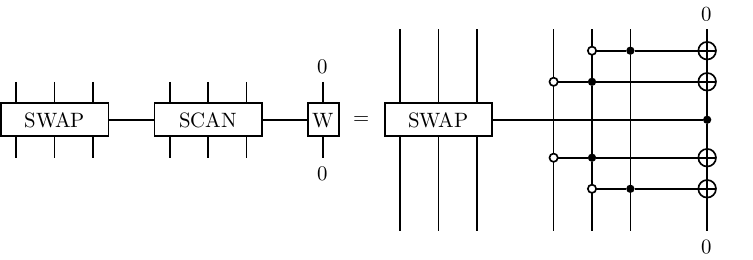}
        \caption{}
        \label{figscan}
    \end{subfigure}
    \begin{subfigure}[hb!]{\textwidth}
        \centering
        \includegraphics[scale=0.75,angle=90]{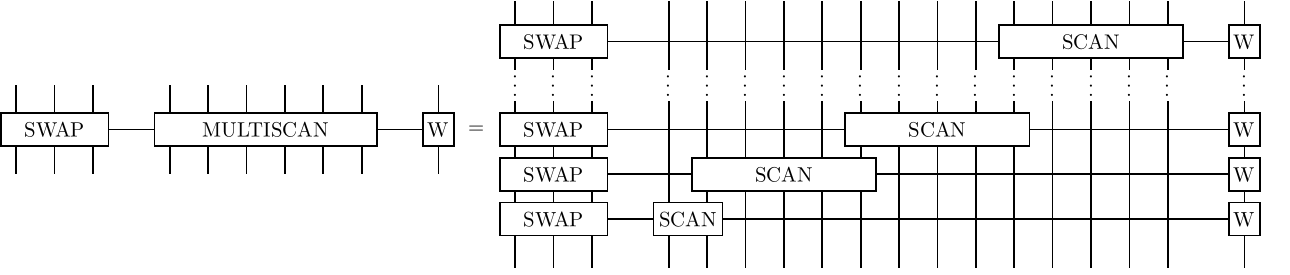}
        \caption{}
        \label{figmultiscan}
    \end{subfigure}
    \caption{Building blocks of the reset circuit in Fig.~\ref{figreset}. a) Search for a magnon as a hole at the third site or a magnon as a particle at the fourth, helped by the information in the first qubit. b) Search and count of domain walls at positions fourth and fifth. The number of qubits of the SCAN gate can vary; here we have chosen it to be three. c) Resource-efficient concatenation of the previous module.}
    \label{figreset_modules}
\end{figure}

\begin{figure}[ht!]
\centering
\includegraphics[scale=0.715,angle=90]{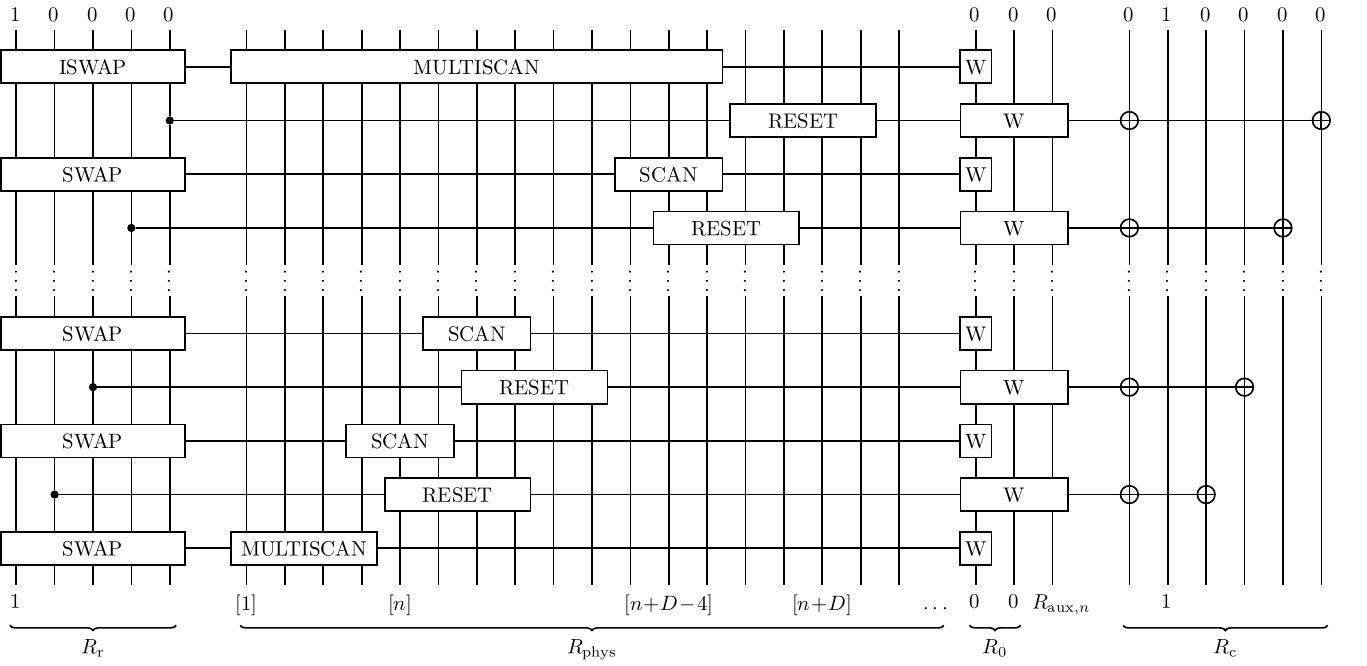}
    \caption{\label{figreset}Circuit to reset the ancillary registers $R_{\mathrm{c}}$ and $R_{\mathrm{r}}$.}
\end{figure}

At this point of the algorithm, the information stored in $R_{\mathrm{c}}$ is also contained in the position of the leftmost magnon in $R_{\mathrm{phys}}$. The circuit in Fig.~\ref{figreset} extracts this information directly from $R_{\mathrm{phys}}$ and uses it to reset $R_{c}$. Starting from the leftmost qubit of $R_{\mathrm{phys}}$, the circuit
searches for  $\ket{01}$
domain walls up to position $n-1$, and stores the result in a new register $R_{\mathrm{r}}$. Since two domain walls of this type cannot be separated by less than four sites, we can save resources as shown in Fig.~\ref{figmultiscan}, by applying the SCAN gate of Fig.~\ref{figscan} to concatenated blocks of five qubits. The first SCAN gate 
acts on the left boundary qubit, which has been omitted, 
together with the first $r+1$ qubits of $R_{\mathrm{phys}}$, where $r$ is the remainder of the quotient $(n-1)/4$.

Let us assume that there was a magnon at $R_{\mathrm{aux},n}$.
The circuit searches for this magnon from position $n+1$ to $n+D$ of $R_{\mathrm{phys}}$. The $n$-th position can be skipped because 
$R_{\mathrm{c}}$ keeps in its starting configuration when the magnon occupies it. First,
the circuit queries whether there is a magnon in the form of a hole at site $n+1$ or a particle at $n+2$. A positive answer is not enough to ensure that the desired qubit has been found. For instance, the configuration  $\ket{0010}$
on sites $(n,\!..,n+3)$
is also compatible with having encountered a previous magnon.
According to the shifting rules~\eqref{displace}, the situation of interest and the spurious one correspond respectively to having one and zero domain walls of type $|01\rangle$ to the left of the magnon.
The module in Fig.~\ref{figreset_w} is able to discriminate between both situations using the information in the register $R_{\mathrm{r}}$. 

The series of concatenated SCAN and RESET gates in Fig.~\ref{figreset} allows to unambiguously identify where the magnon was inserted and correctly reset $R_{\mathrm{0}}$. It is important to notice that the SCAN gates will count now $\ket{01}$ configurations associated both
to domain walls and magnons. This agnostic counting is fundamental for the final reset of the register $R_{\mathrm{r}}$. The design of the circuit is such that a $\ket{01}$ 
count originated by a magnon never leads to an active $R_{\mathrm{r}}$ control qubit in the RESET gates. Finally, following the same scheme of Fig.~\ref{figmultiscan}, a series of SCAN gates is applied from position $n+D-3$ to the first qubit of $R_{\mathrm{phys}}$. They trigger a chain of $\mathrm{CSWAP}$ gates inverse to that in Fig.~\ref{figSWAP}, denoted as $\mathrm{ISWAP}$, ensuring the reset of $R_{\mathrm{r}}$ after their completion.

\begin{table}[ht!]
\centering
\renewcommand{\arraystretch}{1.5}
\setlength{\tabcolsep}{5pt}
\small
\begin{tabular}{c|ccc}
\hline \hline 
 & $\mathrm{CNOT}$ & Toffoli & $\mathrm{CSWAP}$\\
 \hline
Figs.\hspace{2pt}\ref{figDWmod1}--\ref{figDWmod2}  & $(N\!-\!D\!-\!1)(\frac{7}{2}D\!+\!3N\!-\!15)$ & $(N\!-\!D\!-\!1)(5D\!+\!4N\!-\!17)$ & {\color{blue}{$\frac{1}{4}(N\!-\!D)(2D^2\!+\!DN\!-\!7D\!+\!4)$}}\\
 Fig.\hspace{2pt}\ref{figinsert} & $(N\!-\!D)D\!+\!2$   & $(N\!-\!D\!-\!1)(D\!+\!4)\!+\!D\!+\!2$       & $0$\\
 Fig.\hspace{2pt}\ref{figreset} & $4(N\!-\!D\!-\!1)\!+\!2$   & {\color{blue}{$2(N^2\!-\!7N+5)\!+\!D(7N\!-\!9D\!+\!3)$}}  & {\color{blue}{$\frac{1}{8}D(N^2\!+\!5N\!-\!DN\!-\!6D-6)$}}\\
\hline
\end{tabular}
\caption{\label{tab:num_gates} Number of $\mathrm{CNOT}$, Toffoli, and $\mathrm{CSWAP}$ 
required by the circuits constructing a folded XXZ eigenstate with $D$ domain walls out of a purely magnonic state on $N-D$ sites. Values in blue indicate upper bounds.
}
\end{table}

\vspace{5mm}

All the circuits just described are run again, taking now the position $n-1$ of $R_{\mathrm{phys}}$ as reference. We iterate this process until all magnons are depleted from $R_{\mathrm{aux}}$, which ends in the state $|0\rangle^{\otimes N-D}$, while $R_{\mathrm{phys}}$ outputs the desired state $|\Psi_{M,D}\rangle$.
Table \ref{tab:num_gates} shows the number of $\mathrm{CNOT}$, Toffoli and $\mathrm{CSWAP}$ gates required by these circuits. For simplicity, we have considered that all $\mathrm{SWAP}$ gates act on $D/2+1$ qubits, although this overestimates the number of $\mathrm{CSWAP}$s required. For example, the first $\mathrm{SWAP}$ gate in Fig.~\ref{figDWmod1} can be replaced by a single $\mathrm{CSWAP}$ between the second and third qubit of $R_{\mathrm{c}}$. 
Table~\ref{tab:num_gates} gives thus an upper bound to their number. Counting the number of Toffoli gates in the circuit of Fig.~\ref{figreset} leads to a lengthy expression due to the structure of the MULTISCAN gate.
The value listed in Table~\ref{tab:num_gates} is an upper estimate.

\begin{figure}[t!]
\centering
\includegraphics[width=0.5\textwidth]{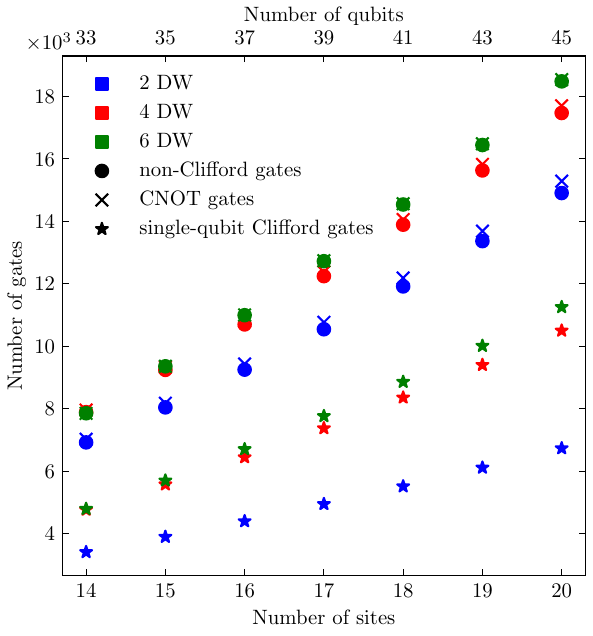}
\caption{\label{fig:gates_clifford} Number of non-Clifford gates, single-qubit Clifford gates, and CNOT gates for $M=2$ magnons and $D=2,\, 4,\, 6$ domain walls as a function of the number of sites. The minimum number of sites is $N=14$, as it is not possible to introduce $D=6$ domain walls in a shorter chain with $M=2$ magnons. The top axis shows the number of qubits in each corresponding circuit.
}
\end{figure}

When analyzing quantum circuits, it is common to differentiate between Clifford and non-Clifford gates. Circuits consisting exclusively of Clifford gates can be efficiently simulated on classical computers~\cite{aaronson_improved_2004} and are relatively easy to implement in a fault-tolerant manner within many quantum error-correcting codes~\cite{raussendorf_fault-tolerant_2007}. However, achieving universal quantum computation requires at least one type of non-Clifford gate.
We decompose our circuit using the gate set $\{R_Z(\theta), R_X(\pi/2), X, \mathrm{CNOT}\}$. The only non-Clifford gates are 
 $R_Z(\theta)$ for $\theta\neq n\pi/2$, where $n$ is an integer.
This gate set is native to IBM quantum computers, making it well-suited for experimental implementation. 
Appendix~\ref{appcompilation} presents the decomposition of the multi-qubit gates used in the circuit into the chosen gate set.
In Figure~\ref{fig:gates_clifford}, we show the number of non-Clifford gates, single-qubit Clifford gates, and CNOT gates for two magnons and varying numbers of domain walls as a function of the number of sites.

\section{Numerical Simulation of the Quantum Algorithm with Noise}

\label{ssim}

In this section, we perform noisy simulations of the circuits preparing simple eigenstates of the folded XXZ model. It is important to stress that the circuit constructed following the instructions in Subsections \ref{ssFreeContact} and \ref{addingDW} is able to prepare eigenstates belonging to any fragment of Hilbert space containing a chosen number of magnons and domain walls. We consider eigenstates
with one magnon and two domain walls in chains with $N=5$
and $N=6$ bulk sites, or equivalently, 7 and 8 sites including the non-dynamical boundary qubits. The reference states of the available fragments are 
\begin{equation}
\begin{split}
\begin{matrix}
N=5: \quad\quad & |10110\rangle \ , \ &|10111\rangle\ , &|10011\rangle\ , & &\\
N=6: \quad\quad &|101100\rangle \ , & |101110\rangle\ , &
|101111\rangle\ , & |100110\rangle\ , &|100111\rangle\ .
\end{matrix}
\end{split}
\label{dw56}
\end{equation}
For the $N=5$ chain, we select the  reference state $|10110\rangle$ and the magnon momentum $p=\pi/4$, while for the $N=6$ chain we choose $|100110\rangle$ and $p=\pi/5$.

\begin{figure}[ht!]
\centering
    \begin{subfigure}[b]{\textwidth}
        \centering
\includegraphics[width=\textwidth]{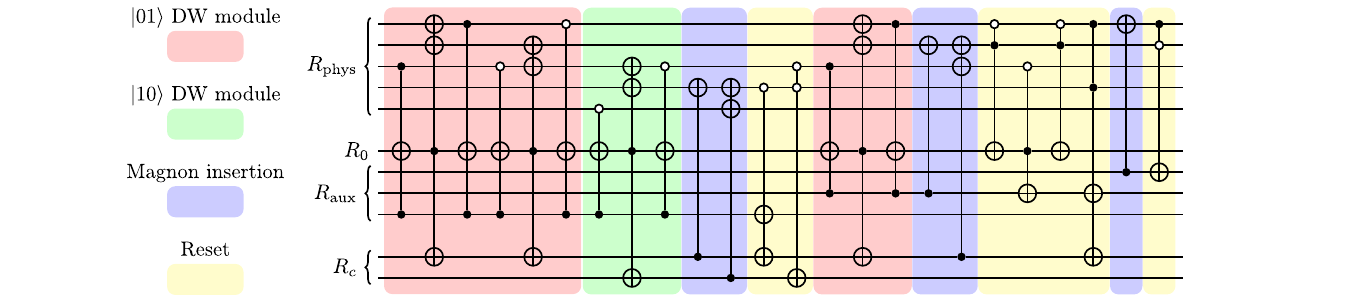}
        \caption{}
        \label{fig5q}
    \end{subfigure}
\begin{subfigure}[b]{\textwidth}
        \centering
\includegraphics[width=\textwidth]{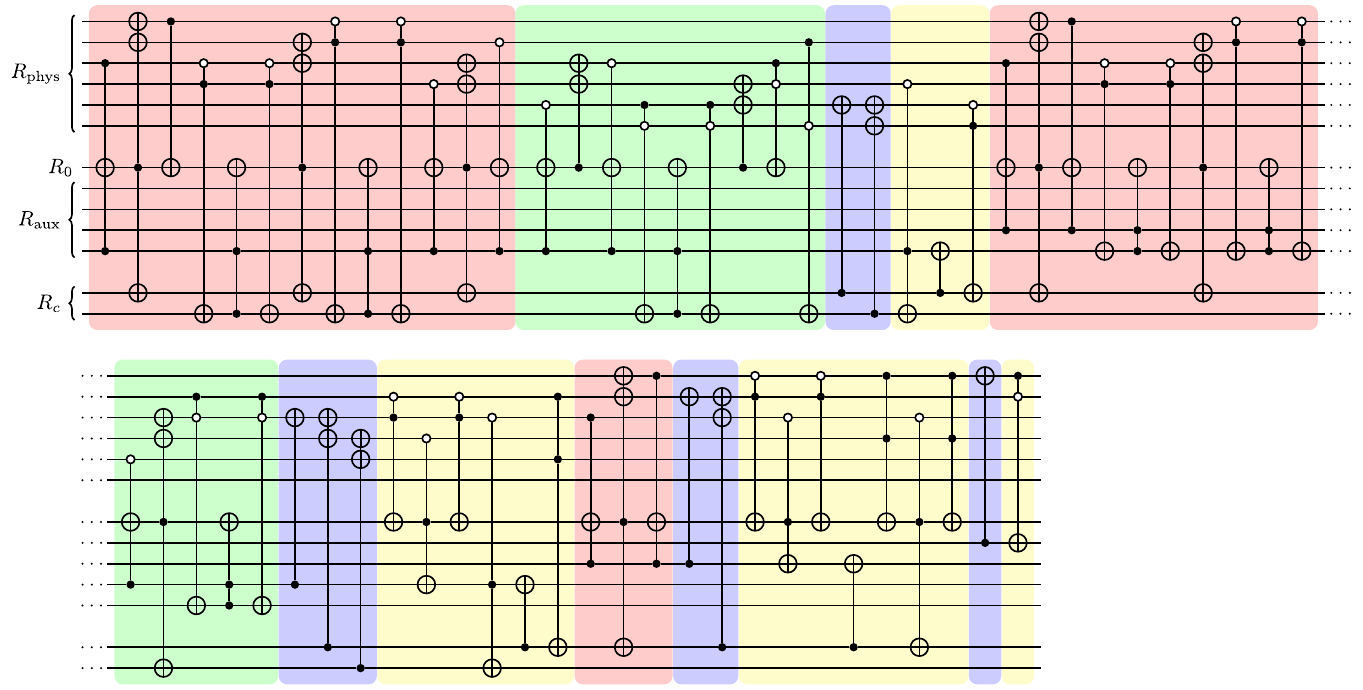} 
        \caption{}
        \label{fig6q}
    \end{subfigure}
    \caption{Circuit representation of $V_D$ used in the numerical simulations. 
    Fig.~\ref{fig5q} and Fig.~\ref{fig6q} correspond to the circuits with 
    $N=5$ and $N=6$, 
    respectively.}
    \label{qc}
\end{figure}

Eigenstates with only two domain walls allow for a simplified version of the counting and reset strategies. Moreover, gates at the beginning or end of the circuits often require less ancilla qubits. In the small chains we consider for simulation, these simplified gates represent are
a large share of the total number. Respecting the general strategy, we have restructured accordingly the circuits to be simulated, shown in Fig. \ref{qc}, in the search for the highest efficiency.
They address first the search for the $|01\rangle$ domain wall, corresponding to the operations shadowed in red, and once completed they turn to the remaining $|10\rangle$ domain wall, with operations shadowed in green. The working register $R_{\mathrm{0}}$ has been reduced to a single qubit, which is reset after each use. For the $N=6$ eigenstate, this reduction requires partially using $R_{\mathrm{c}}$ as workspace. The register $R_{\mathrm{c}}$ counting domain walls can be realized with just two qubits, which switch from $|0\rangle$ to $|1\rangle$ when the first or second domain wall is encountered. This in turns allows to insert the magnon at the appropriate location of the physical register by means of CNOT instead of Toffoli gates, which have been shadowed in blue. Finally, $R_{\mathrm{c}}$ can be reset without the need to introduce an additional register $R_{\mathrm{r}}$, corresponding to the area shadowed in yellow. Our goal is to estimate the circuit's performance on current noisy intermediate-scale quantum (NISQ)~\cite{Preskill18} devices.
The simulations are performed using the Qibo package~\cite{Efthymiou22}. Both the programs and the data necessary to reproduce the results are available in~\cite{code_xxz_folded}. For completeness, Appendix \ref{evolution} describes how the state of $R_{\mathrm{phys}}$ evolves through the circuits \ref{qc}, taking as input the different domain wall reference states in \eqref{dw56}. 

The error rate of two-qubit gates is generally an order of magnitude higher than that of single-qubit gates on the main platforms, typically ranging from $10^{-4}$ to $10^{-3}$~\cite{clark_high-fidelity_2021,ding_high-fidelity_2023,evered_high-fidelity_2023,zhang_tunable_2024}. Therefore, we use a simplified noise model in which only the two-qubit gates are considered noisy.
Coherent noise, which introduces systematic errors, can be transformed into incoherent noise through randomized compiling~\cite{wallman_noise_2016}, as has been experimentally demonstrated on NISQ computers~\cite{hashim_randomized_2021}. A simple model for incoherent noise is depolarizing noise.
The $n$-qubit depolarizing channel is given by
\begin{equation}
    \mathcal{E}(\rho) = (1-\lambda)\rho + \lambda \, \frac{\mathcal{I}}{2^n} \, ,
\end{equation}
where $\lambda$ is the depolarizing rate and $n$ is the number of qubits.
We simulate a noisy two-qubit gate by applying a two-qubit depolarizing channel following the gate operation.
This model characterizes a gate with error rate $r$ if $\lambda=(4r)/3$~\cite{magesan_scalable_2011}.
According to the mentioned error rates, we choose a depolarizing parameter of $3\cdot 10^{-3}$.

Another important aspect to consider is the connectivity of the qubits, which significantly affects the error of two-qubit gates, particularly between distant qubits. Current implementations of low-error gates in superconducting quantum computers utilize static qubits arranged in a two-dimensional grid, which are generally restricted to nearest-neighbor connectivity~\cite{huang_superconducting_2020}. In contrast, trapped-ion setups allow for the shuttling of qubits, enabling them to achieve all-to-all connectivity~\cite{kielpinski_architecture_2002,home_complete_2009,kaushal_shuttling-based_2020}.

\begin{figure}[ht!]
\centering
    \begin{subfigure}[t]{0.42\textwidth}
        \centering
        \includegraphics[width=0.98\textwidth]{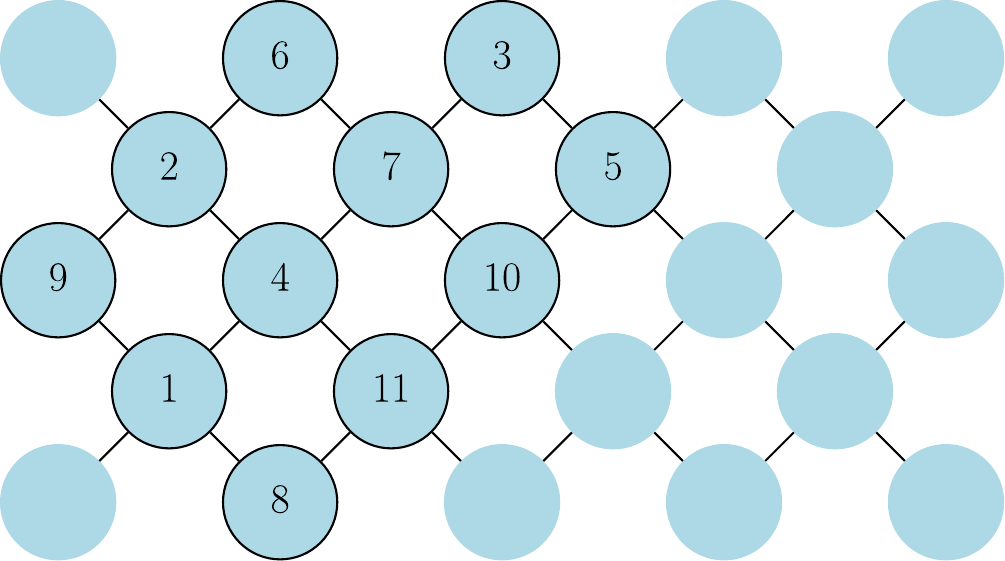}
        \caption{}
        \label{fig_a_conn}
    \end{subfigure} \hspace{.6cm}
    \begin{subfigure}[t]{0.42\textwidth}
    \centering
    \includegraphics[width=0.98\textwidth]{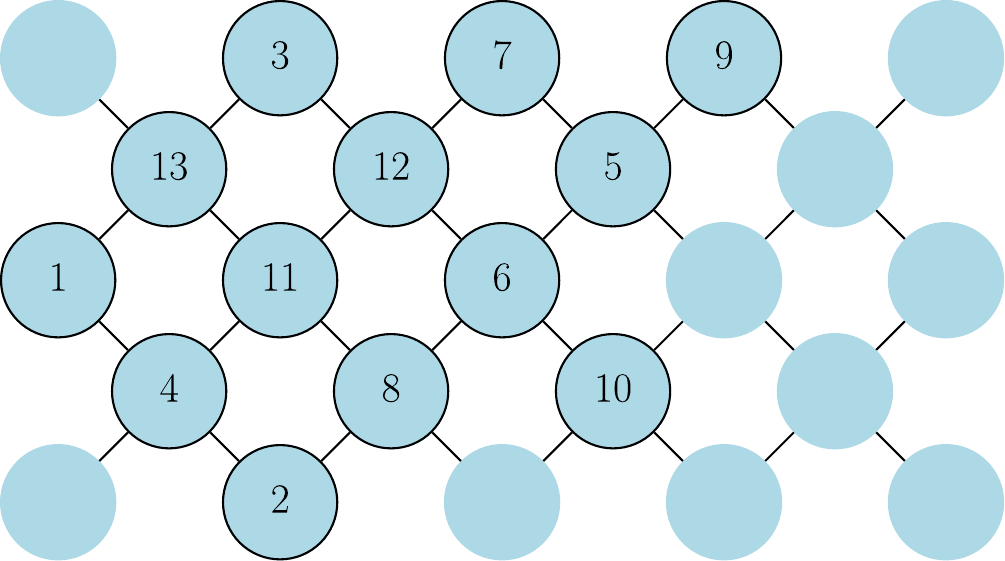}
        \caption{}
        \label{fig_b_conn}
    \end{subfigure}
\caption{\label{fig:qubit_connectivity} Mapping from virtual to physical qubits in the Sycamore23 quantum device. Fig.~\ref{fig_a_conn} corresponds to the state with $N=5$, where 11 physical qubits are used. Fig.~\ref{fig_b_conn} corresponds to the state with $N=6$, where 13 physical qubits are used. Among the 23 available qubits, the qubits we used are highlighted in black. Each number refers to a qubit in our circuit, ordered from top to bottom.}
\end{figure}

This consideration leads us to explore the scenario of all-to-all connectivity, where we decompose the circuit into $\mathrm{CNOT}$ gates and single-qubit gates, permitting connectivity between any pair of qubits. 
For this, we use the gate set $\{R_Z(\theta), R_X(\pi/2), X, \mathrm{CNOT}\}$ mentioned before.
We also consider a scenario with more restricted connectivity, focusing specifically on the architecture of the Google Sycamore23 device~\cite{arute_quantum_2019}, as shown in Fig.~\ref{fig:qubit_connectivity}. To achieve this connectivity, a polynomial number of SWAP gates between connected qubits is necessary.
It is important to find an optimal mapping between the qubits in our circuit and the physical qubits on the quantum device in order to minimize the number of SWAP gates. This problem, known as ``routing'', is solved numerically using the SABRE algorithm~\cite{li_tackling_2019}, which is implemented in Qibo~\cite{Efthymiou22}. The number of $\mathrm{CNOT}$ gates, single-qubit gates, and the circuit depth is given in Table~\ref{tab:num_gates_sim}.

\begin{table}[ht!]
\centering
\renewcommand{\arraystretch}{1.2}
\begin{tabular}{cc|ccccc}
\hline \hline 
& $N$ & $R_Z$ & $R_X(\pi/2)$ & $X$ & $\mathrm{CNOT}$ & Depth\\
\hline
\multirow{ 2}{*}{All-to-all connectivity} &5 & $152$   & $48$  & $14$ & $118$ & $215$\\
&6   &  $392$  & $98$  & $32$ & $316$ & $448$ \\
\hline
\multirow{ 2}{*}{NN connectivity}&5 & $206$   & $98$  & $19$ & $228$ & $361$\\
&6   &  $501$  & $194$  & $42$ & $639$ & $795$ \\
\hline \hline
\end{tabular}
\caption{\label{tab:num_gates_sim} Comparison of the number of quantum gates and circuit depth for all-to-all connectivity and nearest-neighbor (NN) connectivity for the circuits to prepare eigenstates with 5 and 6 bulk qubits.}
\end{table}

To understand the impact of the noise, we simulate the state both with and without noise and calculate the fidelity between the two. The fidelity $F$ between two quantum states $\rho_1$ and $\rho_2$ is defined as
\begin{equation}
F(\rho_1,\rho_2) = \operatorname{tr}\left(\sqrt{\sqrt{\rho_1}\rho_2\sqrt{\rho_1}}\right)^2 \, .
\end{equation}
For the $N=5$ eigenstate, the fidelity is 0.7624 with all-to-all connectivity, but with nearest-neighbor connectivity, it drops to 0.5980.
Similarly, the fidelity of the $N=6$ eigenstate decreases from 0.4841 with all-to-all connectivity to 0.2387 with nearest-neighbor connectivity.
Additionally, we compute the energy of each eigenstate and the expectation values of the charges $Q_1$ and $Q_2$ under both conditions. 
It is important to note that these observables are defined over $N+2$ qubits. Given that the boundary qubits are in the state $|0\rangle$, we project these observables onto this subspace, enabling us to calculate them without simulating the entire states.
The resulting values, along with their relative errors, are shown in Table~\ref{tab:sim_MSE}.

\begin{table}
\centering
\renewcommand{\arraystretch}{1.2}
\begin{tabular}{cc|cccc}
\hline \hline 
&N & $\langle H \rangle$ & $\langle Q_1 \rangle$ & $\langle Q_2 \rangle$ \\
\hline
\multirow{2}{*}{Noiseless}  &5 & $-0.7071$  & $3$ & $4$ \\
           &6 & $-0.8090$  & $3$ & $4$ \\
\hline
\multirow{4}{*}{All-to-all connectivity}   &\multirow{2}{*}{5} 
                    &$-0.5363$ $(0.2415)$ & $2.9085$ $(0.0305)$  & $3.8442$ $(0.0389)$  \\
                    & & $\bf -0.7124$ $\bf (0.0075)$ & $\bf2.9743$ $\bf(0.0086)$  & $\bf3.9969$ $\bf(0.0008)$  \\
&\multirow{2}{*}{6} &$-0.4103$ $(0.4929)$ & $2.9249$ $(0.0250)$  & $3.8078$ $(0.0481)$  \\
                    & &$\bf -0.7786$ $\bf (0.0376)$ & $\bf 2.9753$ $\bf(0.0082)$  & $\bf3.9272$ $\bf(0.0182)$  \\
\hline
                    
\multirow{4}{*}{NN connectivity}     &\multirow{2}{*}{5} 
                    &$-0.4215$ $(0.4039)$ & $2.8409$ $(0.0530)$  & $3.7126$ $(0.0719)$  \\
                    & &$\bf-0.7249$ $\bf(0.0251)$ & $\bf2.9597$ $\bf(0.0134)$  & $\bf3.9426$ $\textbf{}(0.0144)$  \\
&\multirow{2}{*}{6} &$-0.2102$ $(0.7402)$ & $2.8950$ $(0.0350)$  & $3.6791$ $(0.0802)$  \\
                    & &$\bf-0.8209$ $\bf(0.0146)$ & $\bf2.9355$ $\bf(0.0215)$  & $\bf3.8440$ $\bf(0.0390)$  \\
\hline \hline
\end{tabular}
\caption{\label{tab:sim_MSE} Summary of the simulated values for the energy $\langle H \rangle$ and the charges $\langle Q_1 \rangle$, $\langle Q_2 \rangle$, under noiseless and noisy conditions with all-to-all connectivity and nearest-neighbor (NN) connectivity with 5 and 6 bulk qubits.
For each number of qubits and connectivity, the first row displays the values obtained under noise, while the second row shows the corresponding values after error mitigation, highlighted in bold font.
The relative errors with respect to the exact values are shown in parentheses for the noisy cases.}
\end{table}

Error mitigation is central to NISQ devices~\cite{cai_quantum_2023}. Circuits with medium depth, such as ours, are not expected to achieve high fidelity in complex observables without error mitigation.
Clifford circuit data has been leveraged to reduce the impact of noise in quantum computations. Quantum circuits composed mainly of Clifford gates can be efficiently simulated on classical computers~\cite{pashayan_fast_2022}. We utilize Clifford Data Regression (CDR)~\cite{czarnik_error_2021}, an error mitigation technique where near-Clifford circuits are employed to generate a dataset of noisy and exact expectation values for a specific observable. This dataset is then used to train a simple linear model that maps noisy outcomes to their exact counterparts. Once trained, this model is applied to mitigate the noisy expectation values produced by circuits that cannot be efficiently simulated. For further details, refer to the Appendix~\ref{appEM}.
The energy and the expectation values of the charges $Q_1$ and $Q_2$ after mitigation are shown in Table~\ref{tab:sim_MSE}.
Notably, CDR reduces the relative error of each observable by an order of magnitude.
This implies that low raw fidelities do not necessarily render current devices impractical, provided error mitigation is applied.

\section{Applications of the Quantum Algorithm to Other Models}

\label{sQA_other_models}

In this section, we show that our quantum algorithm is not limited to the folded XXZ model, but can also be applied to other problems.

Consider the expression \eqref{wavefunct} for the preparation of the eigenstates of the folded XXZ model. Our key result was the construction of an efficient quantum circuit for the operator $U_D$. 
This operator performs a ``hard-rod deformation'' and it also adds domain walls to the state. The hard-rod deformation can be understood as adding extra length to the particles by replacing every occurrence of a local configuration $\ket{1}$ in the computational basis by $\ket{01}$. To obtain the eigenstates of the folded XXZ model, we applied $U_D$ to an eigenstate of the XX model. However, we can also consider the action of $U_D$ on other states, thereby revealing other applications of our quantum algorithm.

For example, we can act with $U_D$ on eigenstates of the XXZ model. This way we can prepare eigenstates of the so-called ``hard-rod deformed XXZ models'' 
(see \cite{sajat-hardrod}). 
Currently, 
there are no efficient formulas for eigenstate preparation in the XXZ chain; 
nevertheless, eigenstates can be prepared in small volumes or with a small number of magnons, and our algorithm can immediately give the eigenstates of the hard-rod deformed XXZ model as well.

Another application, somewhat independent of the previous computations, is the preparation of a modified version of Dicke states. The Dicke state $\ket{\mathcal{D}_M}$ is defined as the equal weight superposition of the basis states that have $M$ excitations over a ferromagnetic vacuum. By applying our hard-rod deformation algorithm, specialized to zero domain walls, to the states $\ket{\mathcal{D}_M}$, we obtain ``constrained Dicke states''. These states are an equal weight superposition of all basis states that have $M$ excitations such that none of them are on neighbouring sites. Since efficient methods for preparing Dicke states are already known \cite{deterministic-dicke}, our algorithm have access to constrained Dicke states as well.

One application of constrained Dicke sates is in the preparation of a selected zero-energy eigenstate of the celebrated PXP model, derived recently in \cite{crosscap-pxp}. This special eigenstate is built on an equal-weight superposition of double tensor products of constrained Dicke states. Our algorithms can be generalized to prepare the desired final state. This could serve as a compelling first example of an exactly prepared eigenstate for the PXP model.

\section{Discussion}

\label{sconcl}

We now summarize our results and place them in a broader context, highlighting potential applications to benchmarking, quantum advantage, and the study of Hilbert space fragmentation, complementing the discussion of the application of our methods to other models and special states in Section~6.

\subsection{Summary of the Results}

In this paper, 
we constructed a quantum algorithm
capable of preparing
all the eigenstates of the folded XXZ model with open boundaries, 
containing both magnons and domain walls.
To the best of our knowledge,
it is the first efficient quantum algorithm 
that prepares every 
eigenstate of an interacting quantum spin chain. 
Along the way, we extended the reformulation of the Bethe Ansatz in terms of quantum circuits, termed ABC in~\cite{Sopena22,Ruiz23}, from closed to open chains.

The construction of the circuit relies on the description of eigenstates with magnons and domain walls of Section~\ref{sec:eigenstates}, which uses as seed an eigenstate of the XX model on a reduced number of sites. 
We proposed to prepare the XX eigenstate using the open boundary generalization of the ABC.
The magnon contact repulsion is then implemented through the simple coordinate shift \eqref{shift}. Subsequently, the domain walls are displaced from their reference position following the rules \eqref{reloc} and \eqref{displace}, which automatically reveal whether a magnon is to be represented as a particle or a hole.
Even though this picture is equivalent to the effective coordinates of~\cite{Pozsgay21,sajat-hardrod}, it offers a simpler description of the folded XXZ eigenstates which could help to construct eigenstates of other models with similar dynamics.

Our circuits require ancillas, especially relevant to the eigenstates that have domain walls. 
To estimate the performance of circuits on NISQ computers, we ran numerical simulations including a simple depolarizing noise model. Since qubit connectivity has a strong impact on the error of gates between distant qubits, we have considered two different architectures. First, an all-to-all connectivity, typical of \mbox{trapped-ion} and Rydberg-atom prototypes. Second, a square-lattice connectivity, as implemented in Google's superconducting chips. Our simulations indicate that simple eigenstates, containing few magnons and domain walls, 
could be implemented with acceptable fidelity on
quantum computers with two-qubit gate {error} below $10^{-3}$. 

\subsection{Potential Applications}

Hilbert space fragmentation~\cite{fragmentation-scars-review-2,hfrag-review} has attracted much attention in recent years. The reason is that it describes a new type of ergodicity breakdown, hence of violation of the eigenstate thermalization hypothesis. Our results contribute to the possibility of studying this phenomenon on a digital quantum computer.  
For example, it would be interesting to explore the evolution and mixing of the fragments under perturbations {that undo the fragmentation}. 
One could use the fact that our algorithm allows preparing not only eigenstates, but also superpositions of eigenstates from different fragments. These superpositions can be easily implemented by initializing the physical register $R_{\mathrm{phys}}$ of the circuit in Subsection~\ref{addingDW} in a superposition of reference-states $|\Psi_{0,D}\rangle$.
The resulting output could then be fed to a quantum circuit realizing the Trotterized evolution under the desired perturbation. {The Floquet circuits respecting fragmentation described in ~\cite{Langlett21} could serve as a starting point to construct a controlled breakdown of the fragmentation property.}  

As a further potential application, our circuits can be used both for benchmarking and for demonstrating quantum advantage in selected problems. The idea is to measure certain local and non-local correlation functions and compare them to results from classical computations. When compact exact results are available, experimental outcomes could serve as reliable benchmarks. Perhaps more interestingly, our method enables access to quantities for which no closed-form classical results are known, but which could now be explored experimentally on quantum devices.

The folded XXZ model is a recently discovered integrable model, for which there are relatively few exact results available for local or non-local correlation functions. 
One notable exception is the so-called emptiness formation probability in states without domain walls, 
which is analytically known and directly measurable; 
see, for example, Section~XI of~\cite{Pozsgay21}. 
This quantity could serve as a practical benchmark on quantum hardware.
In contrast, 
many other correlation functions, both local and non-local, do not admit compact analytic expressions, 
even though the wavefunctions are known via the Bethe Ansatz. 
This is in fact a very common phenomenon: it is  notoriously difficult to exactly compute non-local correlation functions also in the Heisenberg XXZ model, especially at intermediate distances. 
Although classical computation is in principle possible using the exact Bethe wave functions, 
the evaluation of correlation functions requires summing over an exponentially large number of terms. 
Therefore, 
in the context of the folded XXZ model, 
the combination of efficient eigenstate preparation and quantum measurement of a non-local observable 
would constitute a concrete demonstration of quantum supremacy.

Most of the eigenstates that we treated serve as concrete examples of highly entangled, 
highly non-Clifford, 
and pure and real quantum states that can be efficiently cross-device verified~\cite{Elben20, Carrasco21, Anshu22}. 
While cross-device verification is typically intended for comparing arbitrary quantum states across different quantum computers, 
these particular states provide a compelling test case due to their complex entanglement structure and the difficulty of their classical simulation. 
Additionally, these states are relevant to quantum cryptography, as they allow two remote parties to verify that they share the same quantum state without revealing enough information to identify the state itself, 
aligning with secure quantum verification protocols where minimal information exchange is required~\cite{Denzler25}. 
Thus, 
while the primary motivation for preparing these states lies in their role in quantum many-body physics, 
their structured nature makes them valuable tools for benchmarking quantum devices and advancing cryptographic verification techniques.

\section*{Acknowledgements}

The authors are thankful to Jos{\'e} Carrasco, 
Germ{\'a}n Sierra, Egor Tiunov,
Santiago Varona,
and Hua-Cheng Zhang for stimulating discussions.
The work of E.L., 
R.R., 
and
A.S. 
has been financially supported by the Spanish Agencia Estatal de Investigaci{\'o}n through 
``Instituto de Física Teórica Centro de Excelencia Severo Ochoa CEX2020-001007-S''
and 
PID2021-127726NB-I00 funded by MCIN/AEI/10.13039/501100011033, 
by European Regional Development Fund, 
and the ``Centro Superior de Investigaciones Científicas Research Platform on Quantum Technologies PTI-001'', 
by the Ministerio de Economía, Comercio y Empresa through the Estrategia Nacional de Inteligencia Artificial project call ``Quantum Spain'', 
and by the European Union through the ``Recovery, Transformation and Resilience Plan - NextGenerationEU'' within the framework of the ``Digital Spain 2025 Agenda''.
The work of B.P. was supported by the NKFIH excellence grant TKP2021-NKTA-64.
R. R. is supported by the Universidad Complutense de Madrid, 
Ministerio de Universidades, and the European Union - NextGenerationEU through contract CT18/22.
The authors are grateful to the 
organizers of 
\textit{Exactly Solved Models and Quantum Computing} at Lorenz Center for support and stimulating environment while this work was being completed.
R.R. is grateful to the organizers of 
\textit{Integrable Techniques in Theoretical Physics} at Physikzentrum Bad Honnef for support and
stimulating environment while this work was being completed.

\appendix

\section{Algebraic Bethe Circuits with Open Boundary Conditions}

\label{appOABC}

In this appendix, we present
ABC with open boundary conditions 
of Subsection~\ref{ssOFFABC}. References~\cite{Sopena22,Ruiz23} 
introduced ABC as quantum circuits 
constructed with multi-qubit unitaries that prepare Bethe states 
of spin-$1/2$ chains with
periodic boundary conditions and $U(1)$ symmetry.
We adapt the framework of~\cite{Sopena22,Ruiz23} to open boundary conditions.

It is well-known that the ABA~\cite{Korepin1993kvr,faddeev1996algebraic}
can be formulated in terms of MPS~\cite{Alcaraz03,Alcaraz03ii,Alcaraz06,Katsura10,Murg_2012}, 
the simplest class of one-dimensional tensor networks~\cite{Cirac20}.
The key behind the ABC in~\cite{Ruiz23}
is the representation of Bethe states
in the CBA as translationally
invariant MPS.
The MPS appeared in~\cite{Alcaraz03,Alcaraz03ii,Alcaraz06} (up to the normalization),
but the connection with the CBA remained unnoticed.
As shown in~\cite{Ruiz23}, the Bethe state
of $M$ magnons in a closed spin-$1/2$ chain with $N$ qubits is
\begin{equation}
|\Psi (p_1,\ldots,p_M)\rangle=
\e^{\im {\cal P}_M} 
\sum_{i_j=0,1} \langle 0|^{\otimes M}  \Lambda^{i_N} \ldots \;\Lambda^{i_1}| 1 \rangle^{\otimes M} | i_1 \dots i_{N} \rangle \ ,
\label{eigenCBA}
\end{equation}
where $p_1,\ldots,p_M$ are the momenta of the magnons,  
and $\Lambda$ is the tensor defining the MPS.  
The tensor $\Lambda$ acts on $M$ ancillas,  
producing one physical qubit in the state $|i\rangle$ alongside $M$ ancillas.  
It has one physical index $i$ and two auxiliary indices.  
For fixed $i$, the tensor is
$\Lambda^i$, a $2^M \times 2^M$ matrix in the auxiliary space.  
Moreover,~\cite{Ruiz23} highlighted that,  
if the MPS is initialized in $|n_1\ldots n_r\rangle$ instead of $|1\rangle^{\otimes M}$,  
(\ref{eigenCBA}) prepares Bethe states with $1 \leq r < M$ magnons and momenta $p_{n_1},\ldots,p_{n_r}$. 

To adjust $\Lambda$ to open boundary conditions, 
following Subsection~\ref{ssOFFABC},
we treat incident and reflected plane waves
of every single magnon
independently.
Let us define
the set of
$2M$ momenta
\begin{equation}
    \label{qqmoms}
    q_{2a-1}=p_a\ , \quad q_{2a}^{\phantom{-1}}= -p_{a} \ , \quad a=1,\ldots, M \ .
\end{equation}
Let us also introduce the momenta variables
\begin{equation}
   \label{yvars}
    y_a=\exp(\im q_a) \ .
\end{equation}
The scattering amplitude between the incident and reflected plane waves of $q_a$ with
the plane wave of $q_b$ is $B(q_a,q_b)$. For instance, the scattering amplitude of 
the XXZ model with open boundary conditions is
\begin{equation}
    \label{B}
    B(q_a,q_b)=(1+y_a y_b-2\Delta y_b)(1+y_a^{-1}y_b-2\Delta y_a^{-1})y_b^{-1} \ ,
\end{equation}
where we recall $\Delta$ is the anisotropy. 
The tensor $\Lambda$ of ABC with open boundary conditions
is borrowed from~\cite{Ruiz23}. 
The tensor $\Lambda$ acts on 
$2M$ ancillas and produces 
one physical qubit in the state 
$|i\rangle$ and $2M$ ancillas, and
reads~\footnote{Formula~\eqref{tensorMPS} matches (55) and (57) of~\cite{Ruiz23} if the number of qubits is
$2M$, $x_a$ replace $y_a$, and~$s_{ab}$ replace $B_{ab}$.
}
\begin{equation}
    \label{tensorMPS}
    \begin{split}
        \Lambda^0&=
        \bigotimes_{a=1}^{2M}
        \begin{bmatrix}
            1 & 0 \\
            0 & y_a
        \end{bmatrix}
        \ , \quad
        \Lambda^1=
        \sum_{a=1}^{2M}
    \overset{a-1}{\underset{b=1}{\bigotimes}}
    \begin{bmatrix}
        1 & 0 \\
        0 & -B(q_b,q_a)y_b
    \end{bmatrix}
    \begin{bmatrix}
        0 & 1 \\
        0 & 0
    \end{bmatrix}
    \overset{2M}{\underset{c=a+1}{\bigotimes}}
    \begin{bmatrix}
        1 & 0 \\
        0 & B(q_c,q_a)y_c
    \end{bmatrix}
    \ .
    \end{split}
\end{equation}
Note $\Lambda$ is not unitary. 
We can use~\eqref{eigenCBA} to write
\begin{equation}
\begin{split}
    |\Psi(q_{n_1}\ldots,q_{n_M})\rangle =\e^{\im\mathcal{Q}_a}\sum_{i_j=0,1} \langle 0|^{\otimes 2M}  \Lambda^{i_{N}} \ldots \;\Lambda^{i_1}|n_1\ldots n_M\rangle | i_1 \ldots i_{N} \rangle \\
    =\underset{1\leq \ell_1<\ldots<\ell_{M}\leq {N}}{\sum}
    \vphantom{\left[\sum_{\beta_1,\ldots,\beta_M=1}^{2M} \right]}
    \sum_{\beta_1,\ldots,\beta_{M}=1}^{M}  \epsilon_{\beta_1 \ldots \beta_{M}}
   \left[\underset{1\leq b<a\leq M}{\prod}B(\widetilde{q}_{\beta_a},\widetilde{q}_{\beta_b}) \right]
   \vphantom{\left[\sum_{\beta_1,\ldots,\beta_M=1}^M \right]}
    \left[\vphantom{\underset{1\leq a<b\leq M}{\prod}}
    \!\overset{M}{\underset{a=1}{\prod}} \,\,\widetilde{y}_{\beta_a}^{\,
    \ell_a}\right]\vphantom{\left[\sum_{\beta_1,\ldots,\beta_M=1}^M \right]}
    \vphantom{\left[\sum_{\beta_1,\ldots,\beta_M=1}^M \right]}
    | \ell_1 \ldots \ell_M \rangle \ ,
    \end{split}
\label{partialapp}
\end{equation}
where 
\begin{equation}
    \label{qmomtot}
    \e^{\im\mathcal{Q}_a}=\prod_{a=1}^M\e^{\im q_{n_a}} \ ,
\end{equation}
$1\leq n_1<\ldots < n_M\leq 2M$ and we defined
$\widetilde{q}_{a}:=q_{n_a}$ and $\widetilde{y}_a=y_{n_a}$.
To forbid~\eqref{partialapp} from carrying
momenta $q_{2a}=p_a$ and $q_{2a-1}=-p_a$, we must impose
$n_{2a-1}\neq n_{2a}$.

Bethe states also contain self-scattering amplitudes
if the boundaries are open.
These amplitudes
arise from the interference between the incident and reflected
plane waves of a single magnon.
The self-scattering amplitude with quasi-momentum $q_a$
is $\alpha(q_a)$.
 For instance, 
the self-scattering amplitude of the XXZ model
with open boundary conditions is
\begin{equation}  
    \label{alpha}
   \alpha(q_a)=1-\Delta y_a^{-1} \ .
\end{equation}
The Bethe state of the CBA with open boundary conditions reads
\begin{equation}
    |\Psi_M\rangle= 
    \sum_{\epsilon_a=\pm1}
    \left[\,
    \prod_{b=1}^M \epsilon_b
    \alpha(\epsilon_b p_b)
    \right]
    |\Psi(\epsilon_1 p_1,\ldots, \epsilon_M p_M ) \rangle\ .
    \label{openBethe}
\end{equation}
Self-scattering requires modifying the tensor network compared to the MPS of Bethe states with periodic boundary conditions.  

\begin{figure}[ht!]
\centering
    \begin{subfigure}[t]{0.49\textwidth}
        \centering
        \includegraphics[scale=0.75,angle=90]{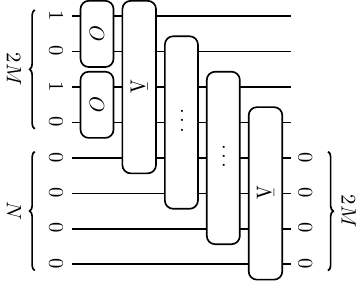}
        \caption{}
        \label{figTN}
    \end{subfigure}
    \begin{subfigure}[t]{0.49\textwidth}
        \centering
        \includegraphics[scale=0.75,angle=90]{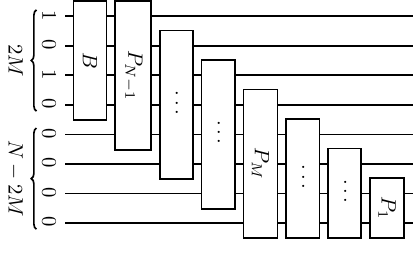}
        \caption{}
        \label{figBOQC}
    \end{subfigure}
    \caption{Realizations of the Bethe Ansatz with open boundary conditions. 
    Fig.~\ref{figTN} depicts the tensor network of~\eqref{openBethe}.
    Fig.~\ref{figBOQC} depicts the ABC to prepare~\eqref{openBethe} with unit norm. }
\end{figure}

We modify the MPS through a boundary operator. 
The boundary operator is a layer of (unnormalized) two-qubit rotations 
of the sector with one qubit at $\ket{1}$
\begin{equation}
    O_a=
    \begin{bmatrix}
        1 & 0 & 0 & 0 \\
        0 & {\alpha}(q_{2a-1})^* & -{\alpha}(q_{2a})& 0 \\
        0 & {\alpha}(q_{2a})^* & {\alpha}(q_{2a-1}) & 0 \\
        0 & 0 & 0 & 1
    \end{bmatrix}
    \ .
\end{equation}
The boundary operator introduces self-scattering amplitudes when the MPS is initialized in the N{\'e}el state like
\begin{equation}
   |10\rangle^{\otimes M} \mapsto \left[\overset{M}{\underset{a=1}{\bigotimes}}\, O_a\right]|10\rangle^{\otimes M}=
   \overset{M}{\underset{a=1}{\bigotimes}}\left[\alpha_{2a-1}|10\rangle - \alpha_{2a}|01\rangle \right]  \ .
\end{equation}
Let us define $\Lambda=\bar{\Lambda}\ket{0}$,  
where $\bar{\Lambda}$ acts on one physical qubit and $2M$ ancillas  
and produces one physical qubit in the state $|i\rangle$ along with $2M$ ancillas.  
Here, $\ket{0}$ represents the input state of the physical qubit.  
The MPS realization of~\eqref{openBethe} forms a staircase of $\bar{\Lambda}$~\cite{Ruiz23}  
that acts on the N{\'e}el state rotated by $B$.  
Fig.~\ref{figTN} depicts the corresponding tensor network.  

The tensor network resembles a circuit acting on  
$N+M$ qubits initialized in $|1\rangle^{\otimes M} |0\rangle^{\otimes N-M}$ and preparing Bethe states.  
However, the equivalence with a quantum circuit does not hold exactly.  
First, the ``gates'' $\bar{\Lambda}$ are not unitary.  
Second, the $M$ rightmost output qubits must be postselected at $|0\rangle$  
to obtain $|\Psi_M\rangle$, leading to an exponential shot cost in $M$.  

Both problems can be addressed by exploiting the large gauge freedom of MPS~\cite{Cirac20}.  
In particular, we can use this local gauge freedom to transform  
the tensor network of Fig.~\ref{figTN} into a quantum circuit.  
Let $X_k$ be the gauge-transformation matrices,
which we choose to preserve the $U(1)$ symmetry of~\eqref{tensorMPS}.
(Closed formulae for $X_k$ and $X_{k}^{-1}$ are
(A2) and (A1) of~\cite{Ruiz23}, respectively).
The action of the gauge transformation on \eqref{tensorMPS} is
\begin{equation}
    \Lambda^{i_k} \mapsto  X_{k+1} \,\Lambda^{i_k} X^{-1}_{k} \ ,
    \label{MPSgauge}
\end{equation}
while the action on the boundary is
\begin{equation}
   \begin{split}
    \overset{M}{\underset{a=1}{\bigotimes}}\left[\alpha_{2a-1}|10\rangle - \alpha_{2a}|01\rangle \right] 
    \mapsto X_1\ \overset{M}{\underset{a=1}{\bigotimes}}\left[\alpha_{2a-1}|10\rangle - \alpha_{2a}|01\rangle \right] \ , \quad
   \langle 0|^{\otimes 2M} \mapsto \langle 0|^{\otimes 2M} X_{N+1}^{-1} \ .
   \end{split}
\end{equation}
Since the gauge transformation preserves the $U(1)$ symmetry,  
the action of $X_{N}^{-1}$ on $\langle 0|^{\otimes M}$ introduces a multiplicative factor.  
The elimination of the ancillas, which can be performed rigorously~\cite{Ruiz24},  
leads to the multi-qubit unitaries $P_k$ of ABC.  
(Closed formulae for $P_k$ follow from $\Lambda$ and $X_k$.)  
Up to this point, our situation and that of~\cite{Ruiz23} are formally identical.  

The action of $X_1$ on the boundary sets our case apart from~\cite{Ruiz23}.  
It requires the ABC with open boundary conditions  
to incorporate a $2M$-qubit unitary $B$ at the boundary,  
ensuring the proper initialization of the staircase of $P_k$  
on the normalized state $|\psi_{\mathrm{in}}\rangle |0\rangle^{\otimes N-2M}$.  
Specifically,~$B$ must accomplish  
\begin{equation}
|\psi_{\mathrm{in}}\rangle=B|10\rangle^{\otimes M}= \rho X_1 \overset{M}{\underset{a=1}{\bigotimes}}\left(\e^{\im q_{2a-1}}\alpha_{2a-1}|10\rangle - \e^{\im q_{2a}}\alpha_{2a}|01\rangle \right) \ ,
\end{equation}
where $\rho$ is the normalization factor ensuring 
$\langle\psi_{\mathrm{in}}|\psi_{\mathrm{in}}\rangle=1$ and
we enclosed the contribution of total 
momentum~(\ref{qmomtot}) 
in the boundary operator.
We do not provide further details on $B$ in general.
Fig.~\ref{figBOQC} depicts ABC with open boundary conditions.

\begin{figure}[ht!]
\centering
    \begin{subfigure}[t]{0.49\textwidth}
        \centering
        \includegraphics[scale=0.75,angle=90]{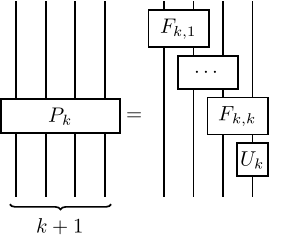}
        \caption{}
        \label{figMGL}
    \end{subfigure}
    \begin{subfigure}[t]{0.49\textwidth}
        \centering
\includegraphics[scale=0.75,angle=90]{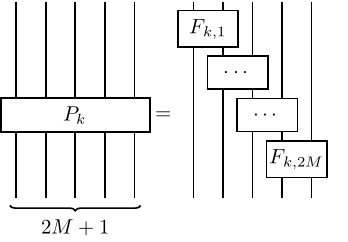}
        \caption{}
        \label{figMGS}
    \end{subfigure}
    \caption{Factorization of $P_k$ into two-qubit unitaries and single-qubit rotations at the \mbox{free-fermion} point.
    Fig.~\ref{figMGL} depicts the factorization of short unitaries $(1\leq k <2M)$.
    Fig.~\ref{figMGS} depicts the factorization of long unitaries $(2M\leq k <N)$.}
\end{figure}

\begin{figure}[hb!]
    \centering
    \includegraphics[scale=0.75,angle=90]{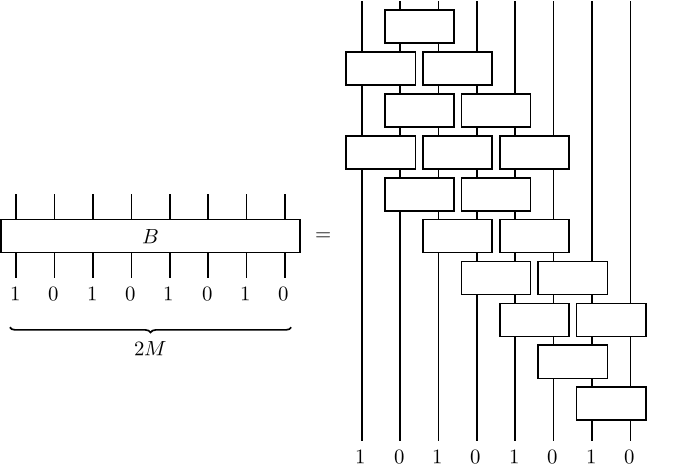}
    \caption{Factorization of the boundary operator $B$ at the free-fermion point. It requires $(M-2)(2M+1$) two-qubit gates if $M-2>0$ and $2M-1$ otherwise.}
    \label{figB}
 \end{figure}

Let us state the connection 
with the quantum circuit for
free fermions with open boundaries of Subsection~\ref{ssOFFABC}. 
We focus on the XXZ model
with open boundary conditions at the free-fermion point $\Delta=0$.
If we set $\Delta=0$ into~\eqref{B} and~\eqref{alpha}, we obtain
\begin{equation}
    B(q_a,q_b)=y_a+\frac{1}{y_a}+y_b+\frac{1}{y_b} \ , \quad \alpha_a=1\ .
\end{equation}
Since
\begin{equation}
    B(p_1,p_2)=B(p_2,p_1) \ , \quad B(p_1,p_2)=B(-p_1,p_2)=B(p_1,-p_2)=B(-p_1,-p_2) \ ,
\end{equation}
we can strip scattering amplitudes out of~\eqref{openBethe} and write
\begin{equation}
    |\Psi_M\rangle=
    \left[\underset{1\leq b<a\leq M}{\prod}B(p_a,p_b) \right]
    \sum_{\epsilon_c=\pm1}
    \left[\,
    \prod_{d=1}^M \epsilon_d
    \right]|\Phi(\epsilon_1 p_1,\ldots, \epsilon_M p_M ) \rangle \ .
    \label{superp}
\end{equation}
Therefore, we can remove the scattering amplitudes as an overall normalization  
of Bethe states with open boundary conditions.  
The result of this removal is~\eqref{superp2}, up to normalization.~\footnote{
If scattering amplitudes were not removed from the outset,  
scattering phases would appear in the multi-qubit unitaries $P_k$.  
These phases are crucial for establishing a connection between the unitaries  
at $\Delta > 0$ and the decomposition into layers of matchgates at $\Delta = 0$.  
Nonetheless, the phases become superfluous if the analysis focuses on the free-fermion point from the outset.  
} 
Moreover, $\ket{\psi_{\mathrm{in}}}$ reduces to
\begin{equation}
    \label{initialXX}
    \ket{\psi_{\mathrm{in}}} = \rho X_1 \left[  \underset{a=1}{\overset{2M}{\bigotimes}}\,{\big(\e^{\im p_a}|10\rangle -\e^{-\im p_a}|01\rangle \big)} \right]  \ ,
\end{equation}
and the unitaries $P_k$ and $B$ decompose into one- and two-qubit gates in Fig.~\ref{figMGL}--\ref{figMGS} and Fig.~\ref{figB}, respectively.~\footnote{
Note that $F_{k,k}$ can be redefined as $F_{k,k}  (1\otimes U_k) \mapsto F_{k,k}$ for $1 \leq k < 2M$ to eliminate the need for \mbox{one-qubit} gates. 
However, 
the resulting two-qubit gate is no longer a matchgate. 
This adjustment can be assumed in the gate counting of Table~\ref{tabmg}, 
effectively replacing the number of matchgates with the number of two-qubit gates.
}
Analytic expressions for the single- and two-qubit unitaries in the decomposition of $P_k$  
can be found in Appendix~\ref{appMG}.  
The numerical optimization that determines the two-qubit gates composing $B$  
is provided in~\cite{code_xxz_folded}.  

\section{Algebraic Bethe Circuits for Free Fermions}

\label{appMG}

The unitaries $P_k$ of ABC for free fermions with open boundaries 
of Subsection~\ref{ssOFFABC}
are layers of two-qubit unitaries, matchgates in particular~\cite{Ruiz23,code_abc},
and a single-qubit rotation for the small unitaries.
Fig.~\ref{figMGL}--\ref{figMGS} shows the factorization of unitaries into matchgates.
In this appendix, we provide closed formulae one- and two-qubit gates that did not appear in~\cite{Ruiz23}.
We also provide a new proof of their unitarity that does not rely on the connection with ABC.

Matchgates are based on the Bethe states
of one magnon with one
quasi-momentum~$q_a$ in $q_1,\ldots,q_{2M}$ over $k$ qubits, where the momenta are~\eqref{qqmoms}.
These Bethe states read
\begin{equation}
    |\Psi_{k,a}\rangle := \overset{k}{\underset{n=1}{\sum}}\,
    y_{a}^{n-1}\,
     \sigma_n^{-}\ket{0}^{\otimes k}
\end{equation}
where $y_a$ are~\eqref{yvars}. 
If $1\leq k <2M$, there are $k$ linearly independent Bethe states.
We choose them to be the Bethe states of momenta $q_1,\ldots,q_k$.
If $2M\leq k <{N}$, there are $M$ linearly independent Bethe states, 
associated to the momenta $q_1,\ldots,q_{2M}$.
The Gram matrix of the Bethe states over $k$ 
qubits thus chosen is positive definite and Hermitian. 
Let
\begin{equation}
    L_k:=\min(k,2M) \ .
\end{equation}
The Gram matrix of the Bethe states reads
\begin{equation}
    \label{Gram}
     C_k =
    \begin{bmatrix}
        C_{11} & C_{12} & \ldots & C_{1L_k} \\
        C_{21} & C_{22} & \ldots & C_{2L_k} \\
        \ldots & \ldots & \ldots & \ldots \\
        C_{L_k1} & C_{L_k2} & \ldots & C_{L_kL_k} \\
    \end{bmatrix}
    \ ,
\end{equation}
where the entries are
\begin{equation}
    \label{Gramentries}
    C_{k,ab} := \langle \Psi_{k,a} | \Psi_{k,b}\rangle \ .
\end{equation}
The matchgates of the quantum circuit are two-qubit unitaries based on the Gram 
matrix:
\begin{equation}
    \label{Fkj}
    F_{k,a}=
    \begin{bmatrix}
        & 1   & 0       & 0 & 0  \\
        & 0   & {u}^*_{k,a} & v_{k,a}  & 0  \\
        & 0   & -{v}^*_{k,a} & u_{k,a}  & 0  \\
        & 0   & 0       & 0  & 1  \\ 
    \end{bmatrix}
    \quad
    1\leq a<L_k \ .
\end{equation}
The entries are
\begin{equation}
    \label{uvmatchgate}
    \begin{split}
        u_{k,a}=\frac{(-1)^{a-1}}{y_1^* \ldots y_{a-1}^*}
        \frac{{\det}_a C_{k+1, a\rightarrow e}}{\sqrt{{\det}_{a-1}C_k {\det}_a C_{k+1}}} \ , 
        \quad v_{k,a}=y_a \sqrt{\frac{{\det}_a C_k{\det}_{a-1}C_{k+1}}{{\det}_{a-1}C_k {\det}_a C_{k+1}}} \ , \\
    \end{split}
\end{equation}
where $\det_a C_k$ denotes the upper-left minor of order  
$a$ of $C_k$ and ${C}_{k,a\rightarrow e}$ 
denotes the matrix that results from the replacement 
of the $a$-th row of ${C}_k$ by 
\begin{equation}
    e=\begin{bmatrix}
        1 &
        1 &
        \dots & 
        1
    \end{bmatrix} .
\end{equation} 
The quantum circuit also comprises the 
single-qubit rotation 
\begin{equation}
    \label{1qU}
    U_k=
    \begin{bmatrix}
        1 & 0 \\
        0 & \exp(\im\phi_k)
    \end{bmatrix}
    \ ,
\end{equation}
where
\begin{equation}
    \label{phase}
    \exp(\im\phi_k)=\prod_{a=1}^{k}\sqrt{\frac{y_{k+1}-y_a}{{y}^*_{k+1}-{y}^*_a}} \ .
\end{equation}
Single-qubit rotations are straightforwardly unitary.
It remains to demonstrate the matchgates~\eqref{Fkj} are unitary as well. 

Unitarity is equivalent to
\begin{equation}
\label{uncond}
|{\det}_a\!\Cc_{k+1,a\to e}\!|^2\!-|y_1\dots y_{a-1}|^2{\det}_{a-1}\Cc_{k} {\det}_a\Cc_{k+1}+|y_1\dots y_a|^2{\det}_a\Cc_{k}{\det}_{a-1}\Cc_{k+1} =0 \ .
\end{equation}
The first step we take is the use of the recursion relation 
\begin{equation}
    \label{recursion}
    C_{k+1,ab}={y}^*_{a}y_{b}C_{k,ab}+1 \ , \quad C_{0,ab}=1 \ ,
\end{equation}
to write
\begin{equation}
    |y_1\dots y_a|^2{\det}_a\Cc_k={\det}_a(\Cc_{k+1}-E) \ ,
\end{equation}
where 
\begin{equation}
    E=
    \begin{bmatrix}
        1 & 1 & \ldots & 1 \\
        1 & 1 & \ldots & 1 \\
        \ldots & \ldots & \ldots & \ldots \\
        1 & 1 & \ldots & 1
    \end{bmatrix} \ .
\end{equation}
Since the subscript $k+1$ is fixed until the end of the proof, we define
\begin{equation}
\Cc:= \Cc_{k+1} \ .
\end{equation}
We use the identity of block-diagonal matrices
\begin{equation}
    \label{identity}
    \det 
    \begin{bmatrix}
        P & Q \\
        R & S
    \end{bmatrix}
    = \det S{\det} (P-RS^{-1}Q) \quad \mathrm{if} \quad \!\!\! \det S \neq 0 \ ,
\end{equation}
to write
\begin{equation}
\label{algids}
\begin{split}
&\frac{{\det}_a\Cc_{a\to e}}{\det_{a-1}\Cc}=1-e^t C_{a-1}^{-1} c_a \ , \\
&\frac{{\det}_a\Cc}{\det_{a-1}\Cc}=C_{aa}-c_a^\dagger C_{a-1}^{-1}c_a \ ,
\\ 
&\frac{{\det}_a(\Cc-E)}{\det_{a-1}(\Cc-E)}=C_{aa}-1-(c_a-e)^\dagger (C_{a-1}-E)^{-1}(c_a-e) \ , 
\end{split}
\end{equation}
where 
\begin{equation}
    c_a = \begin{bmatrix}
        C_{a1} \\
        C_{a2} \\
        \ldots \\
        C_{aa-1}
    \end{bmatrix}
    \ , \quad
    C_a-1 = \begin{bmatrix}
        C_{11} & C_{12} & \ldots & C_{1a-1} \\
        C_{21} & C_{22} & \ldots & C_{2a-1} \\
        \ldots & \ldots & \ldots & \ldots \\
        C_{a-11} & C_{a-12} & \ldots & C_{a-1a-1}
    \end{bmatrix}
    \ .
\end{equation} 
If we introduce~\eqref{algids} into~\eqref{uncond}, we obtain
\begin{equation}
\label{uncond2}
\begin{split}
{\det}_{a-1}\Cc\,[(e^t C_{a-1}^{-1}c_a )(c_a^\dagger C_{a-1}^{-1}e)-e^t C_{a-1}^{-1}c_a-c_a^\dagger C_{a-1}^{-1}e]
+{\det}_{a-1}(\Cc-E) [c_a^\dagger C_{a-1}^{-1}c_a\\
-c_a^\dagger (C_{a-1}-E)^{-1}c_a
+e^t (C_{a-1}-E)^{-1}c_a+c_a^\dagger (C_{a-1}-E)^{-1}c_a
] =0 \ ,
\end{split}
\end{equation}
where we have used that the identity~\eqref{identity} implies
\begin{equation}
{\det}_{a-1}(\Cc-E) [1+e^t (C_{a-1}-E)^{-1}e]={\det}_{a-1}\Cc \ .
\end{equation}
Let us consider now the identity 
\begin{equation}
\label{algids2}
(C_{a-1}-E)^{-1}-E^{-1}=\big(C_{a-1}^{-1}e  \big)\big[e^t(C_{a-1}-E)^{-1}\big] ,
\end{equation}
where we have used the factorization $E=e\otimes e^{t}$. 
The positive definiteness of $C_{a-1}$ enables us to write
\begin{equation}
    1-e^t C_{a-1} e =\det\left[C_{a-1}^{-1/2}(C_{a-1}-E)C_{a-1}^{-1/2}\right] =\frac{\det_{a-1}(C-E)}{\det_{a-1}C} \ .
\end{equation}
The introduction of this identity into~\eqref{algids2} provides
\begin{equation}
    \label{idfinal1}
   \frac{\det_{a-1}(C-E)}{\det_{a-1}C}e^{t}(C_{a-1}-E)^{-1}=e^t C_{a-1}^{-1} \ .
\end{equation}
The positive definiteness of $C_{a-1}-E$ holds due to the recursion relation~\eqref{recursion} and enables us to write
\begin{equation}
    1+e^t (C_{a-1} -E)^{-1}e =\det\left[(C_{a-1}-E)^{-1/2}C_{a-1}(C_{a-1}-E)^{-1/2}\right] =\frac{\det_{a-1}C}{\det_{a-1}(C-E)} \ .
\end{equation}
The introduction of this identity into~\eqref{algids2} provides
\begin{equation}
    \label{idfinal2}
    \frac{\det_{a-1}C}{\det_{a-1}(C-E)}C_{a-1}^{-1}e= (C_{a-1}-E)^{-1}e \ .
\end{equation}
The use of~\eqref{idfinal1} and~\eqref{idfinal2} in~\eqref{uncond2},
makes the left-hand side vanish identically. 
The unitarity of matchgates is thus proven.

\section{Details on the \texorpdfstring{$\boldsymbol{N=5}$}{N=5} and \texorpdfstring{$\boldsymbol{N=6}$}{N=6} Circuits}

\label{evolution}

The circuits in Fig. \ref{qc} build folded XXZ eigenstates with one magnon and two domain walls in chains with $N=5$ and $N=6$ bulk sites. They exhibit some simplifications with respect to the general scheme explained in Subsection \ref{addingDW}. In order to clarify their validity, we detail here how the state of the physical register evolves through the different modules when a magnon is present at the position $n$ of $R_{\mathrm{aux}}$. The register $R_{\mathrm{phys}}$ starts in one of the domain reference states listed in \eqref{dw56}. The position of the $|01\rangle$ domain wall is then updated by the presence of the magnon at $R_{\mathrm{aux},n}$ following the rules in \eqref{reloc} and \eqref{displace}. The position of the $|10\rangle$ domain wall is updated next. These operations correspond respectively to the red- and green-shadowed areas of the circuits in Fig. \ref{qc}. The blue-shadowed areas describe the insertion of the magnon into the appropriate location of $R_{\mathrm{phys}}$ using the information in $R_{\mathrm{c}}$, while the yellow areas reset $R_{\mathrm{c}}$. For convenience, we include below both the state of $R_{\mathrm{phys}}$ and $R_{\mathrm{c}}$ at each stage.

\pagebreak

\begin{center}
    Intermediate states along the $N=5$ circuit
\end{center}
\vspace{3mm}
\small
\begin{equation*}
\renewcommand{\arraystretch}{1.5}
\begin{array}{ccccccccc}
\text{Position of the}\;\;\;\;\; &|\Psi_{0,D}\rangle &&|01\rangle \mathrm{\;DW\; module}&               &|10\rangle \mathrm{\;DW\; module}&               &\!\!\!\!\!\!\mathrm{Magnon \;insertion}\!\!\!\!\!\!&               \\
\text{magnon in}~R_{\mathrm{aux}}  &              &               &                &               &                &               &                & \\
     &|00110\rangle \otimes |00\rangle &\longrightarrow& |11110\rangle \otimes|10\rangle &\longrightarrow& |11000\rangle \otimes |11\rangle &\longrightarrow& |11001\rangle \otimes |11\rangle \\
 n=3\;\;\;\;\;\; &|00111\rangle\otimes |00\rangle &\longrightarrow& |11111\rangle \otimes |10\rangle &\longrightarrow& |11111\rangle \otimes |10\rangle &\longrightarrow& |11101\rangle \otimes |10\rangle \\
     &|00011\rangle \otimes |00\rangle &\longrightarrow& |01111\rangle \otimes |10\rangle &\longrightarrow& |01111\rangle \otimes |10\rangle &\longrightarrow& |01101\rangle \otimes |10\rangle \\
     &               &                &               &             &               &                &               &                \\
     &|00110\rangle \otimes |00\rangle &\longrightarrow& |11110\rangle \otimes |10\rangle &\longrightarrow& |11110\rangle \otimes |10\rangle &\longrightarrow& |11010\rangle \otimes |10\rangle \\
n=2\;\;\;\;\;\;  &|00111\rangle \otimes |00\rangle &\longrightarrow& |11111\rangle \otimes |10\rangle &\longrightarrow& |11111\rangle \otimes |10\rangle &\longrightarrow& |11011\rangle \otimes |10\rangle  \\
     &|00011\rangle \otimes |00\rangle &\longrightarrow& |00011\rangle \otimes |00\rangle &\longrightarrow& |00011\rangle \otimes |00\rangle &\longrightarrow& |01011\rangle \otimes |00\rangle \\
     &               &               &                &               &                &               &                &             \\
     &|00110\rangle \otimes |00\rangle &\longrightarrow& |00110\rangle \otimes |00\rangle &\longrightarrow& |00110\rangle \otimes |00\rangle &\longrightarrow& |10110\rangle \otimes |00\rangle \\
n=1\;\;\;\;\;\;  &|00111\rangle \otimes |00\rangle &\longrightarrow& |00111\rangle \otimes |00\rangle &\longrightarrow& |10111\rangle \otimes |00\rangle &\longrightarrow& |10111\rangle \otimes |00\rangle \\
     &|00110\rangle \otimes |00\rangle &\longrightarrow& |00011\rangle \otimes |00\rangle &\longrightarrow& |00011\rangle \otimes |00\rangle &\longrightarrow& |10011\rangle \otimes |00\rangle \\
\end{array}
\end{equation*}

\pagebreak
\begin{center}
    Intermediate states along the $N=6$ circuit
\end{center}
\vspace{3mm}
\small
\begin{equation*}
\renewcommand{\arraystretch}{1.5}
\begin{array}{ccccccccc}
\text{Position of the}\;\;\;\;\; &|\Psi_{0,D}\rangle &&|01\rangle \mathrm{\;DW\; module}&               &|10\rangle \mathrm{\;DW\; module}&               &\!\!\!\!\!\!\mathrm{Magnon \;insertion}\!\!\!\!\!\!&               \\
\text{magnon in}~R_{\mathrm{aux}}  &              &               &                &               &                &               &                & \\
     &|001100\rangle \otimes |00\rangle &\longrightarrow& |111100\rangle \otimes |10\rangle &\longrightarrow& |110000\rangle \otimes |11\rangle &\longrightarrow& |110001\rangle \otimes |11\rangle \\
     &|001110\rangle \otimes |00\rangle &\longrightarrow& |111110\rangle \otimes |10\rangle &\longrightarrow& |111000\rangle \otimes |11\rangle &\longrightarrow& |111001\rangle \otimes |11\rangle \\
n=4\;\;\;\;\;\;      &|001111\rangle \otimes |00\rangle &\longrightarrow& |111111\rangle \otimes |10\rangle &\longrightarrow& |111111\rangle \otimes |10\rangle &\longrightarrow& |111101\rangle \otimes |10\rangle \\
     &|000110\rangle \otimes |00\rangle &\longrightarrow& |011110\rangle \otimes |10\rangle &\longrightarrow& |011000\rangle \otimes |11\rangle &\longrightarrow& |011001\rangle \otimes |11\rangle \\
     &|000111\rangle \otimes |00\rangle &\longrightarrow& |011111\rangle \otimes |10\rangle &\longrightarrow& |011111\rangle \otimes |10\rangle &\longrightarrow& |011101\rangle \otimes |10\rangle \\
     &|000011\rangle \otimes |00\rangle &\longrightarrow& |001111\rangle \otimes |00\rangle &\longrightarrow& |001111\rangle \otimes |10\rangle &\longrightarrow& |001101\rangle \otimes |10\rangle \\
      &              &               &                &               &                &               &                &                \\
     &|001100\rangle \otimes |00\rangle &\longrightarrow& |111100\rangle \otimes |10\rangle &\longrightarrow& |110000\rangle \otimes |11\rangle &\longrightarrow& |110010\rangle \otimes |11\rangle \\
     &|001110\rangle \otimes |00\rangle &\longrightarrow& |111110\rangle \otimes |10\rangle &\longrightarrow& |111110\rangle \otimes |11\rangle &\longrightarrow& |111010\rangle \otimes |11\rangle \\
n=3\;\;\;\;\;\;      &|001111\rangle \otimes |00\rangle &\longrightarrow& |111111\rangle \otimes |10\rangle &\longrightarrow& |111111\rangle \otimes |10\rangle &\longrightarrow& |111011\rangle \otimes |10\rangle \\
     &|000110\rangle \otimes |00\rangle &\longrightarrow& |011110\rangle \otimes |10\rangle &\longrightarrow& |011110\rangle \otimes |11\rangle &\longrightarrow& |011010\rangle \otimes |11) \\
     &|000111\rangle \otimes |00\rangle &\longrightarrow& |011111\rangle \otimes |10\rangle &\longrightarrow& |011111\rangle \otimes |10\rangle &\longrightarrow& |011011\rangle \otimes |10\rangle \\
     &|000011\rangle \otimes |00\rangle &\longrightarrow& |000011\rangle \otimes |00\rangle &\longrightarrow& |000011\rangle \otimes |10\rangle &\longrightarrow& |001101\rangle \otimes |10\rangle \\
               &               &                &               &                &               &                &                \\
     &|001100\rangle \otimes |00\rangle &\longrightarrow& |111100\rangle \otimes |10\rangle &\longrightarrow& |111100\rangle \otimes |10\rangle &\longrightarrow& |110100\rangle \otimes |10\rangle \\
     &|001110\rangle \otimes |00\rangle &\longrightarrow& |111110\rangle \otimes |10\rangle &\longrightarrow& |111110\rangle \otimes |10\rangle &\longrightarrow& |110110\rangle \otimes |10\rangle \\
n=2\;\;\;\;\;\;      &|001111\rangle \otimes |00\rangle &\longrightarrow& |111111\rangle \otimes |10\rangle &\longrightarrow& |111111\rangle \otimes |10\rangle &\longrightarrow& |110111\rangle \otimes |10\rangle \\
     &|000110\rangle \otimes |00\rangle &\longrightarrow& |000110\rangle \otimes |00\rangle &\longrightarrow& |000110\rangle \otimes |00\rangle &\longrightarrow& |010110\rangle \otimes |00\rangle \\
     &|000111\rangle \otimes |00\rangle &\longrightarrow& |000111\rangle \otimes |00\rangle &\longrightarrow& |000111\rangle \otimes |00\rangle &\longrightarrow& |010111\rangle \otimes |00\rangle \\
     &|000011\rangle \otimes |00\rangle &\longrightarrow& |000011\rangle \otimes |00\rangle &\longrightarrow& |000011\rangle \otimes |00\rangle &\longrightarrow& |010011\rangle \otimes |00\rangle \\
     &               &               &                &               &                &               &                &                \\
     &|001100\rangle \otimes |00\rangle &\longrightarrow& |001100\rangle \otimes |00\rangle &\longrightarrow& |001100\rangle \otimes |00\rangle &\longrightarrow& |101100\rangle \otimes |00\rangle \\
     &|001110\rangle \otimes |00\rangle &\longrightarrow& |001110\rangle \otimes |00\rangle &\longrightarrow& |001110\rangle \otimes |00\rangle &\longrightarrow& |101110\rangle \otimes |00\rangle \\
n=1\;\;\;\;\;\;      &|001111\rangle \otimes |00\rangle &\longrightarrow& |001111\rangle \otimes |00\rangle &\longrightarrow& |001111\rangle \otimes |00\rangle &\longrightarrow& |101111\rangle \otimes |00\rangle \\
     &|000110\rangle \otimes |00\rangle &\longrightarrow& |000110\rangle \otimes |00\rangle &\longrightarrow& |000110\rangle \otimes |00\rangle &\longrightarrow& |100110\rangle \otimes |00\rangle \\
     &|000111\rangle \otimes |00\rangle &\longrightarrow& |000111\rangle \otimes |00\rangle &\longrightarrow& |000111\rangle \otimes |00\rangle &\longrightarrow& |100111\rangle \otimes |00\rangle \\
     &|000011\rangle \otimes |00\rangle &\longrightarrow& |000011\rangle \otimes |00\rangle &\longrightarrow& |000011\rangle \otimes |00\rangle &\longrightarrow& |100011\rangle \otimes |00\rangle \\
\end{array}
\end{equation*}
\normalsize

\section{Error Mitigation Details}

\label{appEM}

\begin{figure}[ht!]
\centering
    \begin{subfigure}[t]{0.49\textwidth}
        \centering
        \includegraphics[width=0.8\textwidth]{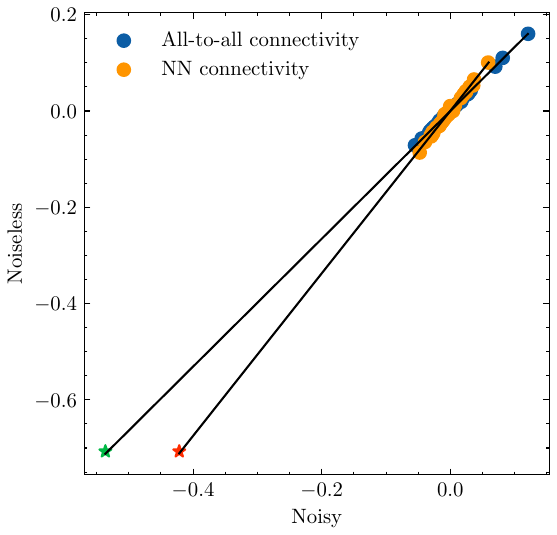}
        \caption{}
        \label{fig_a_mit_e}
    \end{subfigure}
    \begin{subfigure}[t]{0.49\textwidth}
    \centering
    \includegraphics[width=0.8\textwidth]{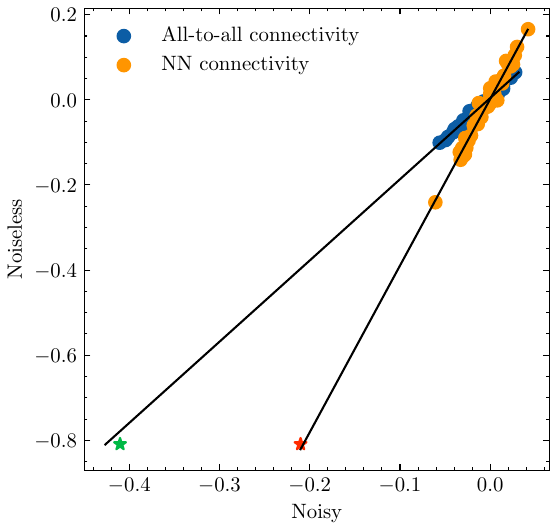}
        \caption{}
        \label{fig_b_mit_e}
    \end{subfigure}
\caption{\label{fig:mitigation_energy} Distribution of exact and noisy energies produced by the near-Clifford training circuits with all-to-all and nearest-neighbor (NN) connectivity. 
Fig.~\ref{fig_a_mit_e} corresponds to the $N=5$ state, while Fig.~\ref{fig_b_mit_e} corresponds to the $N=6$ state.
The green and red stars show the noisy and exact result for the energy for each connectivity.}
\end{figure}

\begin{figure}[ht!]
\centering
    \begin{subfigure}[t]{0.49\textwidth}
        \centering
        \includegraphics[width=0.8\textwidth]{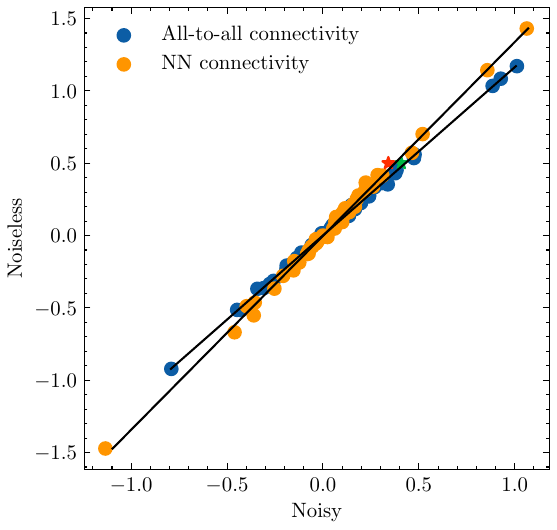}
        \caption{}
        \label{fig_a_mit_q1}
    \end{subfigure}
    \begin{subfigure}[t]{0.49\textwidth}
    \centering
    \includegraphics[width=0.8\textwidth]{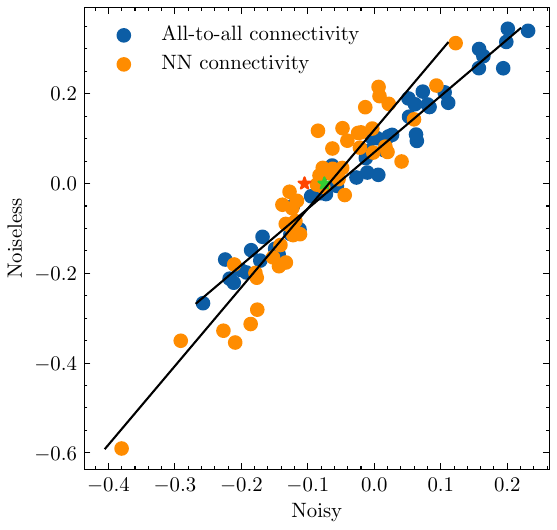}
        \caption{}
        \label{fig_b_mit_q1}
    \end{subfigure}
\caption{\label{fig:mitigation_q1} Distribution of exact and noisy $\langle Q_1\rangle - N/2$ produced by the near-Clifford training circuits with all-to-all and nearest-neighbor (NN) connectivity. 
Fig.~\ref{fig_a_mit_q1} corresponds to the $N=5$ state, while Fig.~\ref{fig_b_mit_q1} corresponds to the $N=6$ state.
The green and red stars show the noisy and exact result for the energy for each connectivity.}
\end{figure}

\begin{figure}[ht!]
\centering
    \begin{subfigure}[t]{0.49\textwidth}
        \centering
        \includegraphics[width=0.8\textwidth]{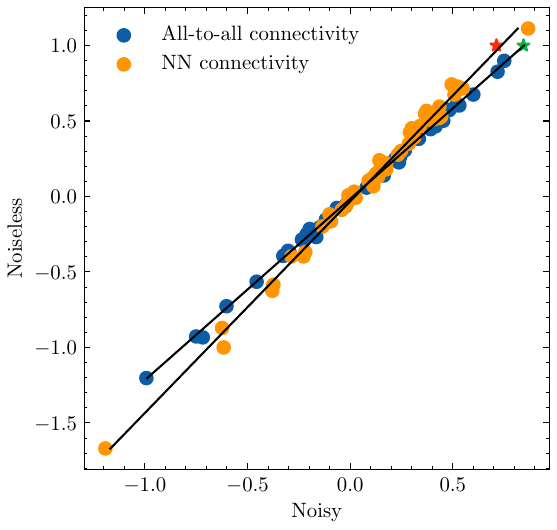}
        \caption{}
        \label{fig_a_mit_q2}
    \end{subfigure}
    \begin{subfigure}[t]{0.49\textwidth}
    \centering
    \includegraphics[width=0.8\textwidth]{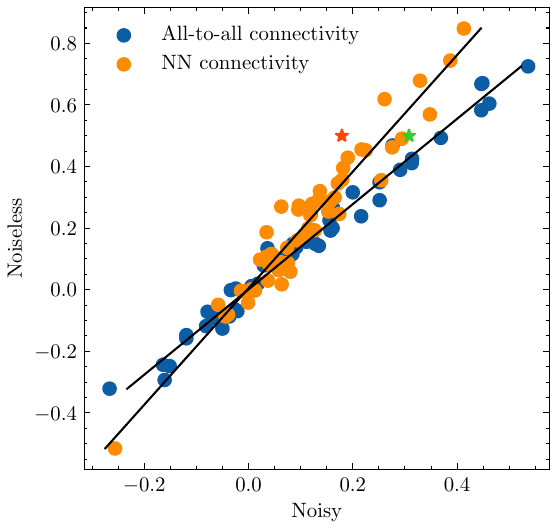}
        \caption{}
        \label{fig_b_mit_q2}
    \end{subfigure}
\caption{\label{fig:mitigation_q2} Distribution of exact and noisy $\langle Q_2\rangle - (N+1)/2$ produced by the near-Clifford training circuits with all-to-all and nearest-neighbor (NN) connectivity. 
Fig.~\ref{fig_a_mit_q2} corresponds to the $N=5$ state, while Fig.~\ref{fig_b_mit_q2} corresponds to the $N=6$ state.
The green and red stars show the noisy and exact result for the energy for each connectivity.}
\end{figure}

The implementation of CDR is very similar to that in~\cite{sopena_simulating_2021}. 
For the training circuits we replace all but 50 non-Clifford gates, selecting them randomly and replacing them probabilistically with a Clifford gate as detailed in~\cite{sopena_simulating_2021}. The exact and noisy expectation values obtained from the training set of near Clifford circuits are used to learn a linear model, which is then applied to the noisy expectation value of the original circuit. We learn a single model to mitigate the energy instead of a different model for each Pauli term. A similar reasoning applies to the charges $\langle Q_1\rangle$ and $\langle Q_2\rangle$, although in this case, we must take into account the constants $N/2$ and $(N+1)/2$ for each charge, respectively. We mitigate these observables without these constants and add them afterward.
An example applying CDR to mitigate the energy with all-to-all and nearest-neighbor connectivity is shown in Fig.~\ref{fig:mitigation_energy}. 
The slope of the lines with nearest-neighbor connectivity are greater, indicating a stronger effective depolarizing noise.
Figs.~\ref{fig:mitigation_q1} and~\ref{fig:mitigation_q2} show similar results for $\langle Q_1\rangle$ and $\langle Q_2\rangle$, respectively

\section{Decomposition of Multi-qubit Gates into Elementary Gates}

\label{appcompilation}

In this appendix, we present the decomposition of the two-qubit gates used as building blocks in the quantum circuits of Sections~\ref{squantalg}–\ref{ssim} into the gate set $\{R_Z(\theta), R_X(\pi/2), X, \mathrm{CNOT}\}$.
This set is used both for gate counting and for simulating the quantum circuits in the presence of noise.

\begin{figure}[ht!]
\centering
\includegraphics[scale=0.65]{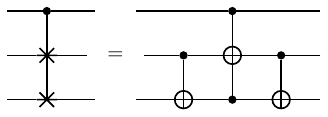}
    \caption{Decomposition of CSWAP gate into elementary gates.}
    \label{figFredkin}
\end{figure}

\begin{figure}[ht!]
\centering
\includegraphics[scale=0.65]{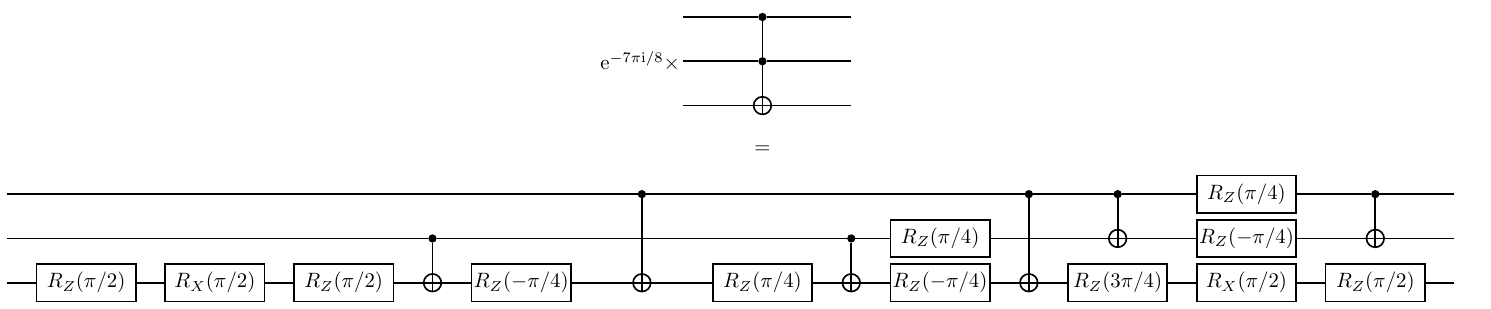}
    \caption{Decomposition of the Toffoli gate into elementary gates, up to an immaterial global phase $\mathrm{e}^{7\pi\mathrm{i}/8}$.}
    \label{figToffoli}
\end{figure}

Figures~\ref{figFredkin} and~\ref{figToffoli} show the decompositions of the CSWAP and Toffoli gates, respectively.

\noindent 

\begin{figure}[ht!]
\centering
\includegraphics[scale=0.65]{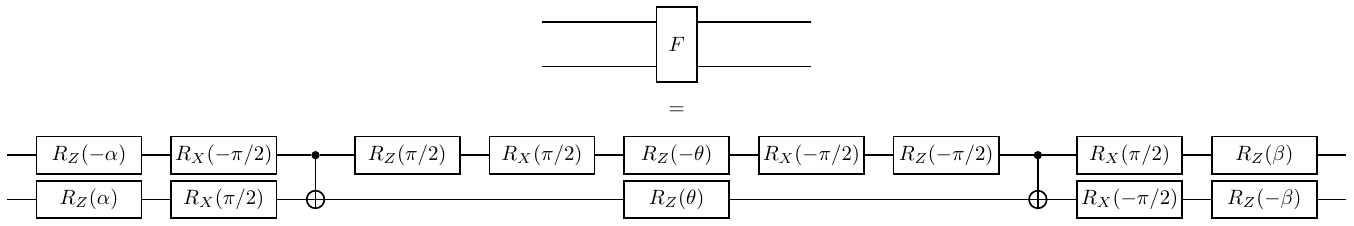}
    \caption{Decomposition of the matchgates \eqref{Fkj} into elementary gates with parameters \eqref{parameters_mg}.}
    \label{figMG}
\end{figure}

Moreover, Fig.~\ref{figMG} shows
the decomposition of the matchgates~\eqref{Fkj}, where the parameters of the decomposition are
\begin{equation}
  \label{parameters_mg}
  u= \e^{\im\varphi}\cos\theta \ , \quad 
  v= \e^{\im\chi}\sin\theta \ , \quad 
  \alpha = \frac{\chi+\varphi-\pi/2}{2} \ , \quad
  \beta = \frac{\chi-\varphi-\pi/2}{2}  \ ,
\end{equation}
and we dropped the subscripts with respect to~\eqref{Fkj} to avoid cluttering the notation. We emphasize that the matchgates obtained by the numerical optimization of Fig.~\ref{figB} are also decomposed according to Fig.~\ref{figMG}.

\bibliographystyle{bibliographicstyle}

\bibliography{references}

\end{document}